\documentclass[journal,twocolumn]{IEEEtran}
\usepackage{amsthm}
\usepackage{lineno}
\usepackage{graphicx}
\usepackage{latexsym}
\usepackage{amsfonts}
\usepackage{subfigure}
\usepackage{dsfont}
\usepackage{epstopdf}
\usepackage{threeparttable}
\usepackage{multirow}
\usepackage{enumerate}
\usepackage{epsfig}
\usepackage{setspace}
\usepackage{caption}
\usepackage{bm}
 \usepackage{color}
\usepackage{cite}
\usepackage{textcomp}
\usepackage{booktabs}
\usepackage{tabu}

\usepackage{makecell}
\def\BibTeX{{\rm B\kern-.05em{\sc i\kern-.025em b}\kern-.08em
    T\kern-.1667em\lower.7ex\hbox{E}\kern-.125emX}}
\usepackage{amsmath}
\usepackage{algorithm}
\usepackage{algpseudocode}

\makeatletter %使\section中的内容左对齐
\renewcommand{\section}{\@startsection{section}{1}{0mm}
  {-\baselineskip}{0.5\baselineskip}{\bf\leftline}}
\makeatother

\usepackage{calligra}
\usepackage{algorithm}
\usepackage{algorithmicx}
\usepackage{algpseudocode}
\usepackage{amsmath}

\usepackage{algorithm}% http://ctan.org/pkg/algorithm

\usepackage{algpseudocode}% http://ctan.org/pkg/algorithmicx

\errorcontextlines\maxdimen

\makeatletter
\newcommand*{\algrule}[1][\algorithmicindent]{\makebox[#1][l]{\hspace*{.5em}\thealgruleextra\vrule height \thealgruleheight depth \thealgruledepth}}%
\newcommand*{\thealgruleextra}{}
\newcommand*{\thealgruleheight}{.75\baselineskip}
\newcommand*{\thealgruledepth}{.25\baselineskip}

\newcount\ALG@printindent@tempcnta
\def\ALG@printindent{%
	\ifnum \theALG@nested>0% is there anything to print
	\ifx\ALG@text\ALG@x@notext% is this an end group without any text?
	\else
	\unskip
	\addvspace{-1pt}% FUDGE to make the rules line up
	\ALG@printindent@tempcnta=1
	\loop
	\algrule[\csname ALG@ind@\the\ALG@printindent@tempcnta\endcsname]%
	\advance \ALG@printindent@tempcnta 1
	\ifnum \ALG@printindent@tempcnta<\numexpr\theALG@nested+1\relax% can't do <=, so add one toRHS and use < instead
	\repeat
	\fi
	\fi
}%
\usepackage{etoolbox}
\patchcmd{\ALG@doentity}{\noindent\hskip\ALG@tlm}{\ALG@printindent}{}{\errmessage{failed to patch}}
\makeatother

\newbox\statebox
\newcommand{\myState}[1]{%
	\setbox\statebox=\vbox{#1}%
	\edef\thealgruleheight{\dimexpr \the\ht\statebox+1pt\relax}%
	\edef\thealgruledepth{\dimexpr \the\dp\statebox+1pt\relax}%
	\ifdim\thealgruleheight<.75\baselineskip
	\def\thealgruleheight{\dimexpr .75\baselineskip+1pt\relax}%
	\fi
	\ifdim\thealgruledepth<.25\baselineskip
	\def\thealgruledepth{\dimexpr .25\baselineskip+1pt\relax}%
	\fi

	\State #1%
	\def\thealgruleheight{\dimexpr .75\baselineskip+1pt\relax}%
	\def\thealgruledepth{\dimexpr .25\baselineskip+1pt\relax}%
}

\usepackage{fancyhdr}

\makeatletter
\newenvironment{breakablealgorithm}
{% \begin{breakablealgorithm}
	\begin{center}
		\refstepcounter{algorithm}% New algorithm
		\hrule height.8pt depth0pt \kern2pt% \@fs@pre for \@fs@ruled
		\renewcommand{\caption}[2][\relax]{% Make a new \caption
			{\raggedright\textbf{\ALG@name~\thealgorithm} ##2\par}%
			\ifx\relax##1\relax % #1 is \relax
			\addcontentsline{loa}{algorithm}{\protect\numberline{\thealgorithm}##2}%
			\else % #1 is not \relax
			\addcontentsline{loa}{algorithm}{\protect\numberline{\thealgorithm}##1}%
			\fi
			\kern2pt\hrule\kern2pt
		}
	}{% \end{breakablealgorithm}
		\kern2pt\hrule\relax% \@fs@post for \@fs@ruled
	\end{center}
}
\makeatother
\begin{document}
%\pagewiselinenumbers
\title{LoS sensing-{based} superimposed CSI feedback for UAV-Assisted mmWave systems}

\author{{Chaojin Qing\textsuperscript{\textbf{a},*},
Qing Ye\textsuperscript{\textbf{a}},
Wenhui Liu\textsuperscript{\textbf{a}},
Zilong Wang\textsuperscript{\textbf{a}},
Jiafan Wang\textsuperscript{\textbf{a}},
Jinliang Chen\textsuperscript{\textbf{b}}}
\\
\bigskip
$^{\text{a}}$ School of Electrical Engineering and Electronic Information, Xihua University, Chengdu, China
\\
$^{\text{b}}$ School of Aeronautics and Astronautics, Xihua University, Chengdu, China
\\

        % <-this % stops a space
\thanks{*Corresponding author. \emph{E-mail address:} qingchj@mail.xhu.edu.cn (C. Qing)}
}

\maketitle
\thispagestyle{fancy}
\fancyhead{}
\cfoot{\quad}                      %清除以前的命令
\lhead{Final Version}
\renewcommand{\headrulewidth}{0pt}

\begin{abstract}
In unmanned aerial vehicle (UAV)-assisted millimeter wave (mmWave) systems, channel state information (CSI) feedback is critical for the selection of modulation schemes, resource management, beamforming, etc.
However, traditional CSI feedback methods lead to significant feedback overhead and energy consumption of the UAV transmitter, therefore shortening the system operation time.
To tackle these issues, inspired by superimposed feedback and integrated sensing and communications (ISAC), a line of sight (LoS) sensing-based superimposed CSI feedback scheme is proposed.
Specifically, on the UAV transmitter side, the ground-to-UAV (G2U) CSI is superimposed on the UAV-to-ground (U2G) data to feed back to the ground base station (gBS).
{At the gBS, the dedicated LoS sensing network (LoS-SenNet) is designed to sense the U2G CSI in LoS and NLoS scenarios.
With the sensed result of LoS-SenNet, the determined G2U CSI from the initial feature extraction will work as the priori information to guide the subsequent operation.
Specifically, for the G2U CSI in NLoS, a CSI recovery network (CSI-RecNet) and superimposed interference cancellation are developed to recover the G2U CSI and U2G data.
As for the LoS scenario, a dedicated LoS aid network (LoS-AidNet) is embedded before the CSI-RecNet and the block of superimposed interference cancellation to highlight the feature of the G2U CSI.}
Compared with other methods of superimposed CSI feedback, simulation results demonstrate that the proposed feedback scheme effectively improves the recovery accuracy of the G2U CSI and U2G data. Besides, against parameter variations, the proposed feedback scheme presents its robustness.


\end{abstract}

\IEEEpeerreviewmaketitle

\section{Introduction}
{Unmanned} aerial vehicle (UAV)-assisted millimeter wave (mmWave) systems have triggered wide research due to their high reliability, excellent flexibility, and large bandwidth availability{\cite{a1}}.
In UAV-assisted mmWave systems, the modulation scheme selection, resource management,  beamforming, etc., require sufficient channel state information (CSI){\cite{a2}}.
{To this end, {the CSI from ground-to-UAV (G2U) links} needs to be estimated by UAV and then fed back to the ground base station ({gBS}) in frequency division duplex (FDD) mode.} {Usually, the gBS in UAV-assisted mmWave systems equips massive antennas to remedy the significant path attenuation of mmWave propagation{\cite{a3}}. This inevitably causes huge feedback overhead at UAVs.} Besides, {prolonging} the battery life of a UAV has always been a challenge to {be considered}.
{For instance, {the limited battery life constraints the operating time of small autonomous rotorcraft and makes it challenging to complete complex exploration and extensive communication missions}{\cite{a4}}}.
{Unfortunately, the necessary CSI feedback substantively increases the energy consumption of the UAV transmitter, which shortens its battery life.}
{However, there is little literature focusing on the G2U CSI feedback for UAV-assisted mmWave systems with low energy consumption, and the existing neural network (NN)-based CSI feedback methods (e.g., \cite{a44,a5,a6}) have not been verified its applicability. Thus, it is vital to develop a CSI feedback scheme to save energy consumption (or prolong battery life).}

As an alternative, the mode of superimposed CSI feedback aims to save battery consumption for transmitting G2U CSI and avoids the extra bandwidth occupation for CSI feedback{\cite{x1}}{\cite{c2}}.
However, {superimposed CSI feedback} inevitably {causes} superimposed interference{\cite{c2}}.
In \cite{x1}, an iterative-based interference cancellation method was proposed, yet with extremely high {computational complexity}.
In \cite{c2}, an extreme learning machine (ELM)-based method was investigated for the superimposed CSI feedback, which improves
the accuracy of CSI recovery and data detection with reduced complexity.
{Nevertheless, the CSI feedback methods in {\cite{x1}}{\cite{c2}} are not dedicated developments for UAV-assisted mmWave systems, so their applicability needs to be further validated.
Especially, the inherent properties of UAV-assisted mmWave systems have not been exploited.
For UAV-to-ground {(U2G)} links, the transmission in line of sight (LoS) scenario (LoS transmission) is usually observed, and the strength of the LoS path {in the LoS scenario }may be critically over 20 dB stronger than those of the non-LoS (NLoS) paths{\cite{a8}}. That is, LoS is a usual scenario (with a high probability) in UAV-assisted mmWave systems{\cite{a7}}, which inspires us to develop LoS features to assist superimposed G2U CSI feedback to alleviate superimposed interference.}

%=================================================FIG1
\begin{figure*}[t]
\centering
\includegraphics[width=5in]{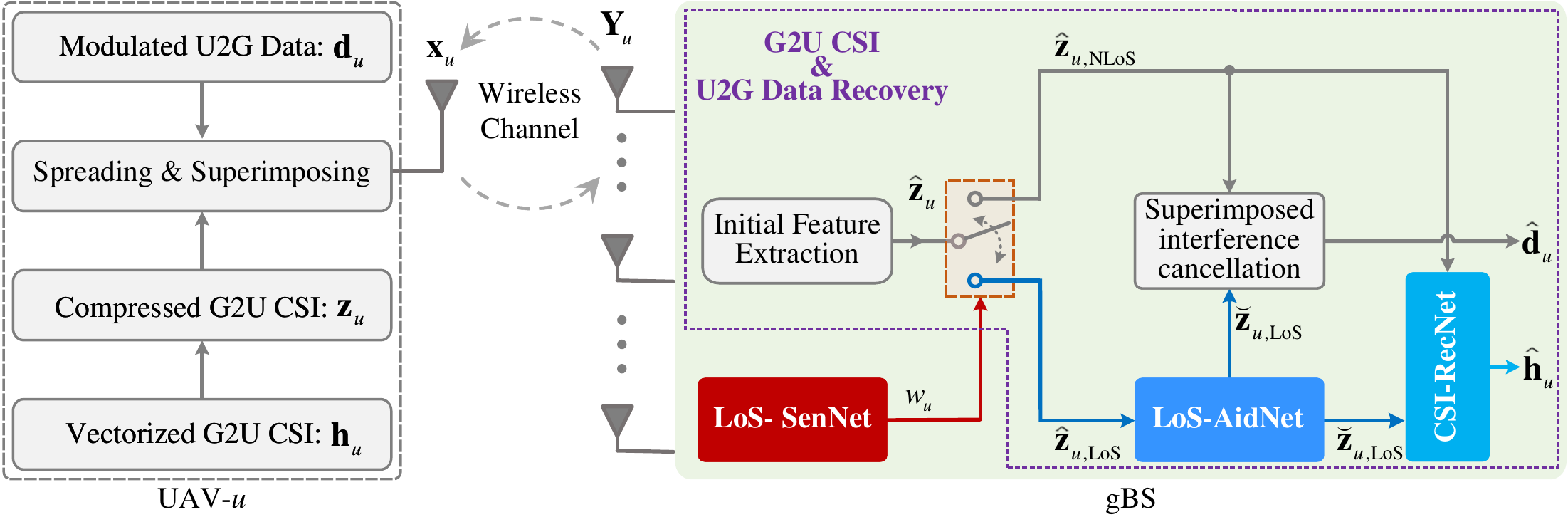}
\caption{System model.\label{fig1}}
\end{figure*}

Recently, integrated sensing and communications (ISAC)-based techniques have aroused great attention, in which sensing signals are derived from received signals for auxiliary communication signals{\cite{i1}}. Many applications have been promoted, e.g., sensing-assisted beam training{\cite{i2}}, sensing-assisted beam tracking and prediction{\cite{i3}}, and sensing-assisted resource allocation{\cite{i4}}, etc. For UAV-assisted mmWave systems, ISAC techniques are developed for the UAV's deployment{\cite{i5}}, the flight trajectory of the UAV{\cite{i6}}, and the transmit beamforming{\cite{i7}}. {Even so, LoS sensing by using the idea of ISAC to assist CSI feedback for UAV-assisted mmWave systems has not been investigated.}
{Inspired by ISAC, the LoS sensing is employed to alleviate superimposed interference of superimposed CSI feedback, {and thus a LoS sensing-assisted superimposed G2U CSI feedback for UAV-assisted mmWave systems is developed in this paper.}}

\subsection*{1.1. Challenge and motivation summary}
The challenges faced in the G2U CSI feedback of UAV-assisted mmWave systems are as follows.
1) Due to the {significant} feedback overhead, the {energy consumption} of UAV transmitters is {significantly} increased, {which hinders UAV prolong its battery life.}
2) {For UAV-assisted mmWave systems, {the inherent property has not yet been exploited that data transmission experiences LoS scenarios with a high probability}}.

Motivated by the {challenges} mentioned above, {this paper} jointly considers the following factors:
1) {It is vital to save energy consumption for UAV-assisted mmWave systems, which promotes us to develop a superimposed mode for its CSI feedback}.
2) Since the LoS {path is} long-lived in UAV scenarios, {it} should be sensed and exploited to alleviate the superimposed interference {of superimposed CSI feedback} and thus improve feedback performance, e.g., the accuracy of the G2U CSI recovery and U2G data detection.
Therefore, we propose a LoS sensing-assisted superimposed CSI feedback scheme {by taking} full advantage of {LoS sensing}.

\subsection*{1.2. Contributions}

To the best of our knowledge, the solution of applying the LoS sensing {to aid} superimposed CSI feedback has not been well studied in the UAV-assisted mmWave system.
The main contributions of this paper are summarized as follows:
\begin{enumerate}
  \item We propose a superimposed CSI feedback scheme for UAV-assisted mmWave systems.
      As far as we know, the superimposed mode for CSI feedback in UAV-assisted mmWave systems has not been well investigated.
      In the proposed scheme, the G2U CSI is superimposed on the U2G data at the UAV transmitter to feed back to the gBS.
      To this end, the energy consumption of the UAV transmitter for CSI feedback is significantly reduced, and the battery life of the UAV is prolonged.
      Besides, the occupation of U2G bandwidth resources is avoided, which improves the spectral efficiency during the CSI feedback phase of UAV-assisted mmWave systems.
      Our work builds a bridge to save the energy consumption of the UAV transmitter and the bandwidth occupation of the UAV-assisted mmWave system, which solves the practical difficulties of CSI feedback for UAV-assisted mmWave systems.

  \item We develop an ISAC-inspired LoS sensing network to alleviate the superimposed interference of the superimposed CSI feedback scheme. To the best of our knowledge, LoS sensing-based superimposed CSI feedback has not been investigated. As the inherent property of U2G links, the LoS scenario occurs with a high probability in UAV-assisted mmWave systems{\cite{a7}}, which is exploited according to the sensing approach to suppress the superimposed interference in this paper.
      Unlike conventional LoS sensing (e.g., \cite{a11} and \cite{a13}), the developed LoS sensing network, named as LoS-SenNet, is inspired by the idea of ISAC. {That is, the same signal, which is received as communication functions in usual systems, is employed for LoS transmission sensing, G2U CSI reconstruction, and U2G data detection at the gBS.} Thus, the LoS sensing-based superimposed CSI feedback scheme is formed without extra hardware/equipment.

  \item We construct lightweight neural networks to reduce the processing latency and computational complexity for the gBS receiver. Due to the single sensing task for LoS transmission, LoS-SenNet is constructed with a lightweight network architecture. Nevertheless, the sensing results from LoS-SenNet are particularly effective to design the lightweight structures of the subsequent networks. With the assistance of LoS-SenNet, LoS aid network (LoS-AidNet) and CSI recovery network (CSI-RecNet) are also constructed with lightweight network architecture. Especially, LoS-AidNet and CSI-RecNet employ the same network architecture, which is beneficial for hardware reuse and thus saves hardware costs. Besides, compared with \cite{x1}, the proposed scheme improves the performance of G2U CSI recovery and U2G data detection with reduced processing latency and computational complexity of the gBS receiver.

\end{enumerate}

\textit{Notations}: Boldface upper case and lower case letters denote matrix and vector, respectively.
${\left(\cdot \right)^T}$ denotes transpose; $\mathbf{I}_P$ is the identity matrix of size $P \times P$;
${\left\|  \cdot  \right\|}$ is the Euclidean norm;
${{\mathrm{Re}}}(\cdot)$ and ${{\mathrm{Im}}}(\cdot)$ represent the operation of taking the real and imaginary parts of a complex value, respectively;
$E[\cdot]$ represents the expectation operation.
$\operatorname{vec}(\cdot)$ denotes the vectorizing of a matrix.

%*******TableI************
\begin{table*}[]

\renewcommand{\arraystretch}{1.2}
\caption{Architecture of LoS-SenNet.}
\label{table_I}
\centering
\scalebox{1}{
\begin{tabu}{@{}c|c|c|c|c@{}}
\tabucline[0.8pt]{-}
{Layer} &  Size of output & Size of kernel & Number of kernel & Activation function \\   \tabucline[0.8pt]{-}
 Input & $L_a\times2N \times 1$ & - & -& - \\ \hline
{{Conv}} & $L_a\times2N \times 1$ & $3\times 3$  & $1$ & ReLU  \\  \hline
{{Maxpool}} &$\lfloor {{L_a}/3} \rfloor  \times \lfloor {N/3} \rfloor  \times 1$ & - & - & -  \\ \hline
{{Flattening}} & $ {\lfloor {{L_a}/3} \rfloor   \lfloor {N/3} \rfloor  }  \times 1$     & -  & - &- \\  \hline
{{Full Connection}} &$1\times1$ & - & - & sigmoid  \\ \tabucline[0.8pt]{-}
\end{tabu}}
\end{table*}

\section{System model}
The system model is given in Fig.~\ref{fig1}, in which one gBS employs a uniform linear array (ULA) with $N$ antennas and $U$ single-antenna UAVs are deployed{\cite{cha1}}. In this section, the channel model and the G2U CSI feedback process are elaborated, respectively.

\subsection*{{2.1. Channel model}}
In a UAV-assisted mmWave system, its wireless channel generally embodies spatial characteristics{\cite{cha1}}.
For G2U links, the channel vector of the $l$th cluster ${{\mathbf{h}}_l} \in {\mathbb{C}^{ 1\times{N}}}$ in the spatial domain can be expressed as{\cite{cha1}}
\begin{equation}
\begin{aligned}
\label{equ: channel model h}
{{\mathbf{h}}_l} = \sum\limits_{k = 1}^K {{\xi _k}} {{\mathbf{a}}_{\text{R}}}\left( {{\theta _{{\text{rx}},k}}} \right){\mathbf{a}}_{\text{T}}^H\left( {{\theta _{{\text{tx}},k}}} \right),
\end{aligned}
\end{equation}
where $K$ is the number of multipath in the $l$th cluster, and ${{\xi _k}}$ is the complex gain of the $k$th multipath in the $l$th cluster.
${{\mathbf{a}}_{\text{R}}}\left( {{\theta _{{\text{rx}},k}}} \right) \in {\mathbb{C}^{{N_r} \times 1}}$ and ${{\mathbf{a}}_{\text{T}}}\left( {{\theta _{{\text{tx}},k}}} \right) \in {\mathbb{C}^{{N_t} \times 1}}$ are the array response vectors for ${{\theta _{{\text{rx}},k}}}$ and ${{\theta _{{\text{tx}},k}}}$ along the $k$th multipath, respectively.
${N_r}$ and ${N_t}$ are the numbers of receiving antennas and transmitting antennas, respectively.
Due to the antenna deployment, we have ${N_r}=1$ and ${N_t}=N$.
Thus, ${{\mathbf{a}}_{\text{T}}}\left( {{\theta _{{\text{tx}},k}}} \right)$ can be written as
${{\mathbf{a}}_{\text{T}}}\left( {{\theta _{{\text{tx}},k}}} \right) = \left[ {1,{e^{ - j2\pi \frac{{d}}{\lambda }\sin \left( \theta  \right)}}, \ldots ,{e^{ - j2\pi \frac{{\left( {{N} - 1} \right){d}}}{\lambda }\sin \left( \theta  \right)}}} \right]$ with ${\lambda }$ and ${d}$ being the G2U link wavelength and the distance between adjacent antennas{\cite{cha2}}, respectively.
Then, on the UAV-$u$ side with $u = 1,2,.\cdots,U,$ the G2U CSI in the time-spatial domain is denoted as
\begin{equation}
\begin{aligned}
\label{equ: channel model H}
\widetilde {\mathbf{H}}_u = {\left[ {{\mathbf{h}}_1^T,{\mathbf{h}}_2^T, \ldots ,{\mathbf{h}}_L^T} \right]^T} \in {\mathbb{C}^{L \times N}},
\end{aligned}
\end{equation}
where $L$ is the number of cluster{\cite{tr1}}.
Subsequently, ${{\mathbf{\widetilde{H}}}_u}$ is transformed into the time-angular domain by {using the inverse discrete Fourier transform}, which is expressed as
\begin{equation}
\begin{aligned}
\label{equ:transmitting_signal_X}
{{\mathbf{\overset{\lower0.5em\hbox{$\smash{\scriptscriptstyle\frown}$}}{H} }}_u} = {{\mathbf{\widetilde H}}_u}{\mathbf{F}_N^H},
\end{aligned}
\end{equation}
where $\mathbf{F}_N$ is an $N \times N$ discrete Fourier transform matrix{\cite{js1}}.
In the time-angular domain, ${{\mathbf{\overset{\lower0.5em\hbox{$\smash{\scriptscriptstyle\frown}$}}{H} }}_u} \in {\mathbb{C}^{L \times N}}$ is typically sparse, and wherein almost all non-zero entries concentrate in the first $L_{a}$ rows of matrix ${{\mathbf{\overset{\lower0.5em\hbox{$\smash{\scriptscriptstyle\frown}$}}{H} }}_u}${\cite{a9}}.
To reduce the feedback overhead, the first $L_{a}$ rows of ${{\mathbf{\overset{\lower0.5em\hbox{$\smash{\scriptscriptstyle\frown}$}}{H} }}_u}$ are extracted and denoted as ${{\mathbf{\overset{\lower0.5em\hbox{$\smash{\scriptscriptstyle\smile}$}}{H} }}_u} \in {\mathbb{C}^{L_{a} \times N}}${\cite{a44}}.
Finally, the vectorized G2U CSI, denoted as ${\mathbf{h}}_u \in {\mathbb{C}^{1 \times L_{a}N}}$, is expressed as
\begin{equation}
\begin{aligned}
\label{equ:transmitting_signal_X}
{{\mathbf{h}}_u} = \operatorname{vec} ( {{{{\mathbf{\overset{\lower0.5em\hbox{$\smash{\scriptscriptstyle\smile}$}}{H} }}}_u}} ).
\end{aligned}
\end{equation}

\subsection*{2.2. G2U CSI feedback}
As shown in Fig.~\ref{fig1}, on the UAV transmitter side, {the G2U CSI is first compressed} by using ${{\mathbf{h}}_u}$ and a random compression matrix $\mathbf{\Phi} \in {\mathbb{C}^{ L_{a}N \times N}}$ to save bandwidth resources and the energy consumption of the UAV transmitter{\cite{y1}}. {Then, the compressed CSI} is spread by the pseudo-random codes (e.g., the Walsh codes{\cite{y1}}) {to alleviate the superimposed interference {caused by} the subsequent process of superimposition}, i.e.,
\begin{equation}
\begin{aligned}
\label{equ:transmitting_signal_X}
\left\{ {\begin{array}{*{20}{l}}
  {{{\mathbf{z}}}_u = {\mathbf{h}}_u \mathbf{\Phi}} \\
  {\mathbf{s}}_u = {\mathbf{{z}}_u}{{\mathbf{Q}}^T}
\end{array}} \right. ,
\end{aligned}
\end{equation}
where ${{\mathbf{z}}}_u \in {\mathbb{C}^{1 \times N}}$ is the compressed G2U CSI, ${\mathbf{s}}_u \in {\mathbb{C}^{1 \times M}}$ is the spread CSI, and $\mathbf{Q} \in {\mathbb{R}^{M \times N}}$ is {the spreading matrix satisfying} ${{\mathbf{Q}}^T}\mathbf{Q} = M{{\mathbf{I}}_N}$.
Here, by superimposing the spread CSI ${\mathbf{s}}_u$ onto the modulated U2G data ${\mathbf{d}}_{u}\in \mathbb{C}^{1\times M}$, the transmitted superimposed signal ${\mathbf{x}}_{u}\in \mathbb{C}^{1\times M}$ of the $u$th UAV is given by{\cite{x1}}
\begin{equation}
\begin{aligned}
\label{equ:transmitting_signal_X}
\mathbf{x}_{u} =& \sqrt{\rho E_{u}}\mathbf{s}_{u} + \sqrt{(1-\rho)E_{u}}\mathbf{d}_{u},
\end{aligned}
\end{equation}
where $\rho \in [0,1]$ stands for the power proportional coefficient of G2U CSI, and $E_{u}$ represents the transmitted power of UAV-$u$.
Without loss of generality, due to the main task of U2G data transmission services, the length of U2G data is longer than that of the compressed G2U CSI, i.e., $M>N$.

At the gBS, after the process of matched-filter, the received signal $\mathbf{Y}_u$ of the UAV-$u$ is given by
\begin{equation}
\begin{aligned}
\label{equ:received_signal_Y}
\mathbf{Y}_{u} =& \mathbf{g}_{u}\mathbf{x}_{u}  + \mathbf{N}_u,
\end{aligned}
\end{equation}
where $\mathbf{N}_u\in \mathbb{C}^{N\times M}$ represents the circularly symmetric complex Gaussian (CSCG) noise with zero-mean and variance ${\sigma_{u}^2}$ for each U2G feedback link, and ${\mathbf{g}}_u \in {\mathbb{C}^{N \times 1}}$ denotes the U2G channel vector from the UAV-$u$ to the gBS.

{To avoid additional hardware devices or reception overhead, inspired by the idea of ISAC, the proposed scheme obtains the signal to be sensed (i.e., U2G channel) and the transmitted superimposed signal (i.e., U2G data and G2U CSI) via the same received signal $\mathbf{Y}_{u}$.}
Specifically, with the received signal $\mathbf{Y}_{u}$, the lightweight LoS-SenNet is developed to sense {the existence} of LoS paths in U2G channels, thereby obtaining the sensed result ${{\chi}_u}$.
Then, by employing the conventional method of superimposed interference cancellation, the initial feature of compressed G2U CSI denoted by ${{\mathbf{\widehat z}}_u}$ is extracted from $\mathbf{Y}_{u}$. With the sensed ${{\chi}_u}$ and extracted ${{\mathbf{\widehat z}}_u}$, the recovery accuracy of the U2G data and G2U CSI are enhanced.

\text{\\}

\section{LoS sensing-based superimposed CSI feedback}
In this section, we present the proposed LoS sensing-based superimposed CSI feedback. {In Section 3.1, the LoS-SenNet will sense the existence of LoS path. With the sensed result, the recovery of the U2G data and G2U CSI is elaborated in Section 3.2.}

\subsection*{3.1. LoS-SenNet}
\fontsize{10pt}{13pt}\selectfont
The LoS-SenNet is developed to exploit the inherent property of UAV-assisted mmWave systems that LoS scenarios appear with a high probability for {U2G} links. {{Inspired by the idea of ISAC{\cite{i1}}, we use the same received signal $\mathbf{Y}_{u}$ to obtain the U2G channel matrix, thereby sensing the existence of LoS paths in U2G channels via LoS-SenNet.}}

\emph{\textbf{Network Design:}}
According to the CNN network strcture in \cite{a11}, {{the LoS-SenNet consists of} a convolutional layer, a maximum pooling layer, a flattening layer, and a fully connected layer.}
The network architecture of LoS-SenNet is summarized in TABLE~\ref{table_I}, and detailed descriptions are given as follows.

Compared with the existing sensing networks (e.g., \cite{a11,a13}), the proposed LoS-SenNet is designed to hold a lightweight network structure. The {input and output sizes} of the convolutional layer are both $L_a\times2N \times 1$. {The size} of the convolution kernel is $3 \times 3$, and the number of convolution kernel is 1.
After the maximum pooling layer and the flattening layer, the output size becomes $ {\lfloor {{L_a}/3} \rfloor   \lfloor {N/3} \rfloor  } \times 1 $.
{Specifically,} the number of neurons of the output layer is $1$ {to obtain the sensed result more intuitively.} {For the activation functions of the convolutional layer and output layer,} the rectified linear unit (ReLU){\cite{a11}} and sigmoid functions are employed, respectively.

{\emph{Remark} 1: The network lightweight of designing LoS-SenNet is mainly embodied in two aspects: 1) only one convolution kernel is designed to capture the features of the LoS path, and 2) only one neuron is employed for the output layer. The main consideration is that the features of the LoS path are easy to {be observed and captured} due to its high strength{{\cite{a8},} and} LoS-SenNet is only used to sense the existence of the LoS path.}

For training the LoS-SenNet, we employ {the} least-squares (LS) channel estimation {for estimating} the U2G channel $\mathbf{g}_{u}$ {to form the U2G channel matrix} ${{\mathbf{\widehat G}}_u} \in \mathbb{C}^{L_a\times N}${\cite{a14}}.
To match the convolutional layer input size and input data type, {the complex-valued ${{\mathbf{\widehat G}}_u}$ is first transformed to a real-valued matrix. Then, the real-valued matrix} is reshaped to ${{\mathbf{\widetilde G}}_u}\in \mathbb{R}^{L_a\times 2N \times1}$, {which is expressed as}
{\begin{equation}
\begin{aligned}
\label{gu}
{{\mathbf{\widetilde G}}_u} = f_\textrm{res}\left( \left [ {\mathrm{Re}({\mathbf{\widehat G}}_u),{\mathrm{Im}}({\mathbf{\widehat G}}_u)} \right ] \right ),
\end{aligned}
\end{equation}}
{where we use {{$f_\textrm{res}\left(  \cdot  \right)$}} to denote} the {reshaping} operation.
Using $\mathbf{\widetilde G}_u$ as the input of the LoS-SenNet, the sensed result $\chi_u$ of whether $\mathbf{\widetilde G}_u$ contains the LoS path is obtained by
\begin{equation}\label{EQ22}
\left\{ \begin{gathered}
  {{{o_u}}} = {f_\textrm{LoS-SenNet}}( {{{{\mathbf{\widetilde G}}}_u},{{\mathbf{\Theta }}_\textrm{LoS-SenNet}}} ) \hfill \\
  {{\chi_u}} = f_\textrm{dec}( {{o_u}} ) \hfill \\
\end{gathered}  \right.,
\end{equation}
where ${o _u}$ is the output of the LoS-SenNet, ${f_{\textrm{LoS-SenNet}}}\left(  \cdot  \right)$ denotes {the mapping} of LoS sensing operation, and ${{\mathbf{\Theta }}_\textrm{LoS-SenNet}}$ is the {network parameter.} The {$f_{\textrm{dec}}(\cdot)$ }is the hard decision operation {with a threshold of 0.5}, yielding two types of ${{\chi}_u}$, i.e., 0 and 1. ${{\chi}_u} = 1$ indicates the U2G channel is with LoS path, otherwise, ${{\chi}_u} = 0$.

%*******TableI************
\begin{table*}[]
\renewcommand{\arraystretch}{1.2}
\caption{Architecture of LoS-AidNet and CSI-RecNet.}
\label{table_II}
\centering
\scalebox{1}{
\begin{tabu}{@{}c|c|c|c|c|c|c@{}}
\tabucline[0.8pt]{-}
\multirow{2}{*}{Layer} & \multicolumn{2}{c|}{Input} & \multicolumn{2}{c|}{Hidden} & \multicolumn{2}{c}{Output} \\ \cline{2-7}
 & \multicolumn{1}{c|}{LoS-AidNet} & \multicolumn{1}{c|}{CSI-RecNet} & \multicolumn{1}{c|}{LoS-AidNet} & \multicolumn{1}{c|}{CSI-RecNet} & \multicolumn{1}{c|}{LoS-AidNet} & \multicolumn{1}{c}{CSI-RecNet} \\ \tabucline[0.8pt]{-}
Neuron number       & $2N$    & $2N$     & $4N$  & $4L_aN$    & $2N$     & $2L_aN$    \\ \hline
Activation function & None & None & LReLU & tanh & Linear & Linear  \\ \tabucline[0.8pt]{-}
\end{tabu}}
\end{table*}

\subsection*{3.2. Superimposed CSI recovery}
During {the sensing phase}, we utilize {the superimposed} interference cancellation method in \cite{x1} to obtain the initial feature of the {compressed} G2U CSI {in} LoS or NLoS {scenario}.
Subsequently{, with} the sensed result ${{\chi}_u}$ and the initial feature of the compressed G2U CSI, {the processing to recover} G2U CSI and U2G data {is performed}.

\subsubsection{Initial feature extraction}
To obtain the initial feature of {compressed} G2U CSI from the received signal $\mathbf{Y}_{u}$, we adopt {the superimposed} interference cancellation method according to \cite{x1}.
Specifically, as in \cite{x1}, we first perform a despread operation on the received signal $\mathbf{Y}_{u}$ to obtain a despread signal ${{\mathbf{V}}_u} \in {\mathbb{C}^{N \times N}}$, i.e.,
\begin{equation}
\begin{aligned}
\label{ds}
{{\mathbf{V}}_u} = {{{{\mathbf{Y}}_u}{\mathbf{Q}}} \mathord{\left/
 {\vphantom {{{{\mathbf{Y}}_u}{\mathbf{Q}}} M}} \right.
 \kern-\nulldelimiterspace} M}.
\end{aligned}
\end{equation}
{Then, the minimum mean squared error (MMSE) estimation of G2U CSI is obtained according to ${{\mathbf{V}}_u}$ [8], which is expressed by
\begin{equation}
\begin{aligned}
\label{csimmse}
{{\mathbf{\overset{\lower0.05em\hbox{$\smash{\scriptscriptstyle\frown}$}}{z} }}_u} = {f_{{\textrm{MMSE}}}}( {{{\mathbf{V}}_u}} ),
\end{aligned}
\nonumber
\end{equation}
where ${{\mathbf{\overset{\lower0.05em\hbox{$\smash{\scriptscriptstyle\frown}$}}{z} }}_u}\in {\mathbb{C}^{1 \times N}}$ is the estimated G2U CSI and ${f_{{\textrm{MMSE}}}}(  \cdot  )$ denotes the mapping function of the MMSE estimator.}
With ${{\mathbf{\overset{\lower0.05em\hbox{$\smash{\scriptscriptstyle\frown}$}}{z} }}_u}$, the interference cancellation technique is utilized to eliminate the impact of G2U CSI on the detection of U2G data{\cite{x1}}, i.e.,

\begin{equation}
\begin{aligned}
\label{ddd}
{{\mathbf{\overset{\lower0.05em\hbox{$\smash{\scriptscriptstyle\frown}$}}{Y} }}_u} = {{\mathbf{Y}}_u} - \sqrt {{{\rho {E_u}} \mathord{\left/
 {\vphantom {{\rho {E_u}} N}} \right.
 \kern-\nulldelimiterspace} N}} {{\mathbf{g}}_u}{{\mathbf{\overset{\lower0.05em\hbox{$\smash{\scriptscriptstyle\frown}$}}{z} }}_u}{{\mathbf{Q}}^T}.
\end{aligned}
\end{equation}
{Subsequently, the MMSE detection is used to obtain the initial detected U2G data ${{\mathbf{\overset{\lower0.05em\hbox{$\smash{\scriptscriptstyle\frown}$}}{d} }}_u}\in {\mathbb{C}^{1 \times M}}$, i.e., ${{\mathbf{\overset{\lower0.05em\hbox{$\smash{\scriptscriptstyle\frown}$}}{d} }}_u} = {\mathcal{D}_{{\textrm{MMSE}}}}( {{\mathbf{\overset{\lower0.05em\hbox{$\smash{\scriptscriptstyle\frown}$}}{Y} }}_u} )$ with ${\mathcal{D}_{{\textrm{MMSE}}}}(  \cdot  )$ denoting the mapping function of the MMSE detector.}
With the initial detected U2G data ${{\mathbf{\overset{\lower0.05em\hbox{$\smash{\scriptscriptstyle\frown}$}}{d} }}_u}$, the impact of U2G data on the estimated G2U CSI is eliminated by utilizing the interference cancellation technique, i.e., ${{\mathbf{\overset{\lower0.05em\hbox{$\smash{\scriptscriptstyle\smile}$}}{Y} }}_u} = {{\mathbf{Y}}_u} - \sqrt {\left( {1 - \rho } \right){E_u}} {{\mathbf{g}}_u}{{\mathbf{\overset{\lower0.5em\hbox{$\smash{\scriptscriptstyle\frown}$}}{d} }}_u}$.
{Similar to (\ref{ds}), we utilize ${{\mathbf{\overset{\lower0.05em\hbox{$\smash{\scriptscriptstyle\smile}$}}{Y} }}_u}$ to obtain an improved despread signal ${{\mathbf{\overset{\lower0.5em\hbox{$\smash{\scriptscriptstyle\smile}$}}{V} }}_u}$, i.e., ${{\mathbf{\overset{\lower0.5em\hbox{$\smash{\scriptscriptstyle\smile}$}}{V} }}_u} = {{{{\mathbf{\overset{\lower0.05em\hbox{$\smash{\scriptscriptstyle\smile}$}}{Y} }}_u}{\mathbf{Q}}} \mathord{\left/
 {\vphantom {{{{\mathbf{Y}}_u}{\mathbf{Q}}} M}} \right.
 \kern-\nulldelimiterspace} M}$.
Finally, we perform MMSE estimation on ${{\mathbf{\overset{\lower0.5em\hbox{$\smash{\scriptscriptstyle\smile}$}}{V} }}_u}$ to obtain the initial feature of compressed G2U CSI ${{\mathbf{\widehat z}}_u} \in {\mathbb{C}^{1 \times N}}$, i.e.,
\begin{equation}
\begin{aligned}
\label{zzz}
{{\mathbf{\widehat z}}_u} = {f_{{\textrm{MMSE}}}}( {{\mathbf{\overset{\lower0.5em\hbox{$\smash{\scriptscriptstyle\smile}$}}{V} }}_u} ).
\end{aligned}
\end{equation}}

{It should be noted that, the MMSE detection is employed to present the initial feature extraction to hold the same detection method in \cite{x1}. Other detection methods, e.g., {zero} forcing (ZF) detection, can also be adopted.}

\subsubsection{G2U CSI and U2G data recovery}
After initial feature extraction, we perform the recovery of G2U CSI and U2G data.
{Specifically, with the sensed result ${{\chi}_u}$, the initial feature of {compressed} G2U CSI ${{\mathbf{\widehat z}}_u}$ with LoS or NLoS is determined due to the same propagation environment{\cite{a12}}}. For the convenience of description, we denote the {compressed} G2U CSI with LoS as ${{\mathbf{\widehat z}}_{u{\text{,LoS}}}}$ and the {compressed} G2U CSI without LoS as ${{\mathbf{\widehat z}}_{u{\textrm{,NLoS}}}}$.
 {Then, for ${{\mathbf{\widehat z}}_{u{\textrm{,NLoS}}}}$, we develop the CSI recovery network (CSI-RecNet) and superimposed interference cancellation to recover G2U CSI and U2G data.
For ${{\mathbf{\widehat z}}_{u{\textrm{,LoS}}}}$, relative to the G2U CSI in NLoS, a dedicated LoS aid network (LoS-AidNet) is embedded {(between }the CSI-RecNet and the block of superimposed interference cancellation) to highlight the feature of compressed G2U CSI.}

Due to the {significant} feature of the G2U CSI {in LoS scenarios{\cite{a13}}, a lightweight network architecture {for LoS-AidNet} is adopted. {For the CSI-RecNet}, the assistance of the extracted {features} by using {the initial feature extraction or LoS-AidNet} is exploited. {With the extracted features}, the CSI-RecNet can learn along with {${{\mathbf{\widehat z}}_{u{\textrm{,NLoS}}}}$ or the output of LoS-AidNet}, and thus can be designed as a lightweight network as well.
To this end, both LoS-AidNet and CSI-RecNet are designed as the same neural network structure containing only a single hidden layer, which is beneficial for hardware reuse to save its costs.
The network architecture of LoS-AidNet and CSI-RecNet are summarized in TABLE~\ref{table_II}, and the detailed descriptions are given as follows.

In LoS-AidNet, {the neurons} of the input layer, hidden layer, and output layer {are set as} $2N$, $4N$, and $2N$, respectively. {For CSI-RecNet, $2N$, $4L_aN$, and $2L_aN$ neurons are adopted for the input layer, hidden layer, and output layer, respectively.}
The leaky rectified linear unit (LReLU) and tanh{\cite{a9}} are employed as activation functions for the hidden {layers} of LoS-AidNet and CSI-RecNet, respectively, and linear activation is adopted as {the activation functions} of their output {layers}.
For training the LoS-AidNet, we first convert {complex-valued} ${{\mathbf{\widehat z}}_{u{\textrm{,LoS}}}} \in {\mathbb{C}^{1 \times N}}$ {to real-valued} ${{\mathbf{\widetilde z}}_{u{\textrm{,LoS}}}} \in {\mathbb{R}^{1 \times 2N}}$ {according to}

\begin{equation}\label{EQ21}
{{\mathbf{\widetilde z}}_{u{\textrm{,LoS}}}} = [ {\mathrm{Re}( {{{{\mathbf{\widehat z}}}_{u{\textrm{,LoS}}}}} ),\mathrm{Im}( {{{{\mathbf{\widehat z}}}_{u{\textrm{,LoS}}}}} )} ].
\end{equation}
{Then, the LoS-AidNet is used to refine the compressed G2U CSI by exploiting the sensed LoS scenario, which is expressed as}
\begin{equation}\label{EQ22}
{{\mathbf{\overset{\lower0.05em\hbox{$\smash{\scriptscriptstyle\smile}$}}{z} }}_{u{\textrm{,LoS}}}} = f_{\textrm{LoS-AidNet}}({{\mathbf{\widetilde z}}_{u{\textrm{,LoS}}}}, \mathbf{\Theta}_{\textrm{LoS-AidNet}}),
\end{equation}
where $f_{\textrm{LoS-AidNet}}(\cdot)$ and $\mathbf{\Theta}_{\textrm{LoS-AidNet}}$ denote {the mapping} of LoS learning operation and its network parameter, respectively.
{{According to} (\ref{ddd}) along with ${{\mathbf{\overset{\lower0.05em\hbox{$\smash{\scriptscriptstyle\smile}$}}{z} }}_{u{\text{,LoS}}}}$,} the U2G data ${{\mathbf{\widehat d}}_u}$ {is detected}.

{The sensed LoS and NLoS scenarios share the same network CSI-RecNet.} By denoting the real-valued form of ${{{{\mathbf{\widehat z}}}_{u{\textrm{,NLoS}}}}}$ as ${{\mathbf{\overset{\lower0.05em\hbox{$\smash{\scriptscriptstyle\smile}$}}{z} }}_{u,{\textrm{NLoS}}}} \in {\mathbb{R}^{1 \times 2N}}$, we have
\begin{equation}\label{EQ23}
{{\mathbf{\overset{\lower0.05em\hbox{$\smash{\scriptscriptstyle\smile}$}}{z} }}_{u,{\textrm{NLoS}}}} = \left[ {\mathrm{Re}\left( {{{{\mathbf{\widehat z}}}_{u{\textrm{,NLoS}}}}} \right),\mathrm{Im}\left( {{{{\mathbf{\widehat z}}}_{u{\textrm{,NLoS}}}}} \right)} \right] .
\end{equation}
Then, using the CSI-RecNet, the recovered G2U CSI ${{{\mathbf{\widehat h}}}_u}$ {is given by}
\begin{equation}\label{EQ24}
{{{\mathbf{\widehat h}}}_u} = {f_{{\textrm{CSI-RecNet}}}}( {{{{\mathbf{\overset{\lower0.05em\hbox{$\smash{\scriptscriptstyle\smile}$}}{z} }}}_u},{\mathbf{\Theta} _{{\textrm{CSI-RecNet}}}}} ).
\end{equation}
where ${{\mathbf{\overset{\lower0.05em\hbox{$\smash{\scriptscriptstyle\smile}$}}{z} }}_u}$ denotes the input of CSI-RecNet,{ i.e.,} ${{\mathbf{\overset{\lower0.05em\hbox{$\smash{\scriptscriptstyle\smile}$}}{z} }}_{u,{\textrm{NLoS}}}}$ or ${{\mathbf{\overset{\lower0.05em\hbox{$\smash{\scriptscriptstyle\smile}$}}{z} }}_{u{\textrm{,LoS}}}}$. $f_{\textrm{CSI-RecNet}}(\cdot)$ and $\mathbf{\Theta}_{\textrm{CSI-RecNet}}$ denote {the mapping} of G2U CSI recovery operation and its network parameter, respectively.

\subsection*{{3.3. Off-line training and online deployment}}

The details of the dataset collection are given as follows.
{Due to the effects in the UAV scenarios (geographical and geomorphic differences, weather effects, etc.), the completeness of the collected real data cannot be guaranteed. Besides, the complete collection of real data is costly and time-consuming{\cite{realdata}}. Thus, we employ the Clustered-Delay-Line (CDL) channel model of 5G standard{\cite{tr1}} to capture the spatial characteristics in the UAV-assisted mmWave system. According to 3GPP TS 38.901{\cite{tr1}}, the CDL channel model is widely used for link level evaluation, which is considered to be very close to the real scenarios{\cite{cdlreal}}.
Specifically, the CDL model is often implemented by phase initialization along four different polarizations and generating coefficients for each cluster{\cite{tr2}}.
We employ the CDL-A and CDL-D models to generate the G2U and U2G channels.
The CDL-A is used to {form} NLoS transmission channels, {and} the CDL-D is used to {generate} LoS transmission channels{\cite{tr1}}.
The proposed LoS sensing-based superimposed CSI feedback scheme is summarized in Algorithm \ref{algorithm1}.

\subsubsection*{{1) Off-line training}}
{The training set $\{ {\widetilde {\mathbf{G}}_u^{\left( {tr} \right)},{e_u}} \}$ is used to train the LoS-SenNet}, where ${{{e}}_u} \in {\left\{ {0,1} \right\}}$ is the training {label} of LoS-SenNet. {${{{e}}_u}=1$ and ${{{e}}_u}=0$ are respectively employed} to label the U2G channel with LoS path or not.
The training data $\widetilde {\mathbf{G}}_u^{\left( {tr} \right)}$ is formed according to (\ref{gu}).
For the {LoS-AidNet}, the training data is formed according to (\ref{EQ21}), {and thus forms the training set $\{ {{\mathbf{\widetilde z}}_{u{\textrm{,LoS}}}}^{\left( {tr} \right)}, {{\mathbf{ z}}_{u}} \}$.}
{According to (\ref{EQ22}) and (\ref{EQ23})}, we collect ${\mathbf{\overset{\lower0.05em\hbox{$\smash{\scriptscriptstyle\smile}$}}{z} }}_{u,{\text{LoS}}}^{\left( {tr} \right)}$ and ${\mathbf{\overset{\lower0.05em\hbox{$\smash{\scriptscriptstyle\smile}$}}{z} }}_{u,N{\text{LoS}}}^{\left( {tr} \right)}$ {to form training set $\{ {\mathbf{\overset{\lower0.05em\hbox{$\smash{\scriptscriptstyle\smile}$}}{z} }}_{u}^{\left( {tr} \right)}, {{\mathbf{h}}_{u}} \}$ to} train the CSI-RecNet.
{The training sets of LoS-SenNet, LoS-AidNet, and CSI-RecNet have $100,000$, $60,000$, and $30,000$ samples, respectively, while the validation sets of them have $10,000$, $6,000$, and $3,000$ samples, respectively.}

In addition, the equivalent signal-to-noise ratio (SNR) in decibel (dB) is defined as $\text{SNR} = 10\textrm{log}_{10}(E_u/\sigma^2_u)$ according to \cite{y1}.
The LoS-SenNet is trained in a noise-free setting. The training SNR of LoS-AidNet is set as 10dB.
When training the CSI-RecNet, the training SNR is set as 20dB.
The Adam optimizer is used to update the network parameters of each network{\cite{a151}}.
We utilize $\alpha$, $\gamma$, and $\xi$ to represent the training epochs of LoS-SenNet, LoS-AidNet, and CSI-RecNet, respectively, where $\alpha=20$, $\gamma=50$, and $\xi=50$.}

The optimization goal of LoS-SenNet is to minimize the mean squared error (MSE) between ${o_u}$ and ${{{e}}_u}$, which is {expressed} as
\begin{equation}\label{LoS-SenNet}
\begin{gathered}
\mathop {\min }\limits_{{\mathbf{\Theta} _{{\textrm{LoS-SenNet}}}}} E[ {{{\| {{f_{\textrm{LoS-SenNet}}}( { \widetilde {\mathbf{G}}_u^{\left( {tr} \right)} ,{{\mathbf{\Theta }}_{\textrm{LoS-SenNet}}}} ) - {{{e}}_u}} \|}^2}} ].
\end{gathered}
\end{equation}
{For training LoS-AidNet, the MSE of} the {refined compressed} G2U CSI with LoS path, i.e., $E[ {{{\| {{{\mathbf{\overset{\lower0.05em\hbox{$\smash{\scriptscriptstyle\smile}$}}{z} }}_{u,{\text{LoS}}}^{\left( {tr} \right)}- {{\mathbf{ z}}_{u}}}} \|}^2}} ]$ {is minimized}, which is given by
\begin{equation}\label{LoS-AidNet}
\begin{gathered}
\mathop {\min }\limits_{{\mathbf{\Theta} _{{\textrm{LoS-AidNet}}}}} E[ {{{\| {f_{\textrm{LoS-AidNet}}({{\mathbf{\widetilde z}}_{u{\text{,LoS}}}}^{\left( {tr} \right)}, \mathbf{\Theta}_{\textrm{LoS-AidNet}}) - {{\mathbf{ z}}_{u}}} \|}^2}} ].
\end{gathered}
\end{equation}
Similarly, the CSI-RecNet minimizes the MSE of the recovered G2U CSI, i.e., $E[ {{{\| {{{{{\mathbf{\widehat h}}}_u}^{\left( {tr} \right)} - {{\mathbf{h}}_{u}}}} \|}^2}} ]$, which {is expressed as}
\begin{equation}\label{CSI-RecNet}
\begin{gathered}
\mathop {\min }\limits_{{\mathbf{\Theta} _{{\textrm{CSI-RecNet}}}}} E[ {{{\| {{f_{{\textrm{CSI-RecNet}}}}( {{\mathbf{\overset{\lower0.05em\hbox{$\smash{\scriptscriptstyle\smile}$}}{z} }}_{u}^{\left( {tr} \right)},{\mathbf{\Theta} _{{\textrm{CSI-RecNet}}}}} ) - {{\mathbf{h}}_{u}}} \|}^2}} ].
\end{gathered}
\end{equation}
We perform the training once for LoS-SenNet, LoS-AidNet, and CSI-RecNet, and save the trained network parameters for the {online running}.

\begin{breakablealgorithm}

	\caption{The algorithm of the proposed LoS sensing-
        based superimposed CSI feedback scheme.}
	\label{algorithm1}
	\begin{algorithmic}[0]

  %      \Statex

    \State{\textbf{[Off-line training stage]}:}
    \State{\textbf{Input:}} Training sets: $\{ {\widetilde {\mathbf{G}}_u^{\left( {tr} \right)},{e_u}} \}$, $\{ {{\mathbf{\widetilde z}}_{u{\textrm{,LoS}}}}^{\left( {tr} \right)}, {{\mathbf{ z}}_{u}} \}$, $\{ {\mathbf{\overset{\lower0.05em\hbox{$\smash{\scriptscriptstyle\smile}$}}{z} }}_{u}^{\left( {tr} \right)}, {{\mathbf{h}}_{u}} \}$; the training epochs of the LoS-SenNet, LoS-AidNet, and CSI-RecNet: $\alpha$, $\gamma$, $\xi$.
%            \end{spacing}
    \State{\textbf{Output:}} The network parameters: ${\mathbf{\Theta }}_{{\text{LoS-SenNet}}}$, ${\mathbf{\Theta }}_{{\text{LoS-AidNet}}}$,\\${\mathbf{\Theta }}_{{\text{CSI-RecNet}}}$.
    \begin{spacing}{0.5}
    \end{spacing}
    \State {LoS-SenNet:}
        \State {Randomly initialize the parameter of LoS-SenNet as ${\mathbf{\Theta }}_{{\text{LoS-SenNet}}}$}.
%		\Function{xxx}{xxx} \label{xxx1}
		\For{$i = 1,2, \ldots \alpha $}
		\myState{Update the network parameter ${\mathbf{\Theta }}_{{\text{LoS-SenNet}}}$ by }
        \State {using the Adam optimizer and (\ref{LoS-SenNet}) with current }
        \State {dataset $\{ {\widetilde {\mathbf{G}}_u^{\left( {tr} \right)},{e_u}} \}$}.
		\EndFor
        \State {Save the network parameter ${\mathbf{\Theta }}_{{\text{LoS-SenNet}}}$}.
%		\EndFunction \label{xxx2}

%        \Statex
        \State {LoS-AidNet:}
        \State {Randomly initialize the parameter of LoS-AidNet as ${\mathbf{\Theta }}_{{\text{LoS-AidNet}}}$}.
%		\Function{xxx}{xxx} \label{xxx1}
		\For{$i = 1,2, \ldots \gamma$}
		\myState{Update the network parameter ${\mathbf{\Theta }}_{{\text{LoS-AidNet}}}$ by }
        \State {using the Adam optimizer and (\ref{LoS-AidNet}) with current }
        \State {dataset $\{ {{\mathbf{\widetilde z}}_{u{\textrm{,LoS}}}}^{\left( {tr} \right)}, {{\mathbf{ z}}_{u}} \}$}.
		\EndFor
        \State {Save the network parameter ${\mathbf{\Theta }}_{{\text{LoS-AidNet}}}$}.
%		\EndFunction \label{xxx2}

%        \Statex
        \State {CSI-RecNet:}
        \State {Randomly initialize the parameter of CSI-RecNet as ${\mathbf{\Theta }}_{{\text{CSI-RecNet}}}$}.
%		\Function{xxx}{xxx} \label{xxx1}
		\For{$i = 1,2, \ldots \xi $}
		\myState{Update the network parameter ${\mathbf{\Theta }}_{{\text{CSI-RecNet}}}$ by }
        \State {using the Adam optimizer and (\ref{CSI-RecNet}) with current }
        \State {dataset $\{ {\mathbf{\overset{\lower0.05em\hbox{$\smash{\scriptscriptstyle\smile}$}}{z} }}_{u}^{\left( {tr} \right)}, {{\mathbf{h}}_{u}} \}$}.
		\EndFor
        \State {Save the network parameter ${\mathbf{\Theta }}_{{\text{CSI-RecNet}}}$}.
%		\EndFunction \label{xxx2}
		%算法中的空行
        \Statex
        \State{\textbf{[Online running stage]:}}
        \State{\textbf{Input:}} The received signal $\widetilde {\mathbf{Y}}_{u}$.
        \State{\textbf{Output:}} The detected U2G data $\widehat {\mathbf{d}}_u^{\left( {te} \right)}$ and the recovered G2U CSI ${\mathbf{\widehat h}}_u^{\left( {te} \right)}$.
        \begin{spacing}{0.5}
        \end{spacing}
        \State {Perform LS channel estimation on $\widetilde {\mathbf{Y}}_{u}$ to gain the U2G channel matrix $\widetilde {\mathbf{G}}_u^{\left( {te} \right)}$.}
        \State {Obtain the sensed result $\chi_u^{\left( {te} \right)}$ via loading $\widetilde {\mathbf{G}}_u^{\left( {te} \right)}$ and the network parameter ${\mathbf{\Theta }}_{{\text{LoS-SenNet}}}$ into LoS-SenNet and then performing the hard decision operation}.
        \State {Perform the initial feature extraction to obtain the compressed G2U CSI initial feature ${{\mathbf{\widehat z}}_u}^{\left( {te} \right)}$ from $\widetilde {\mathbf{Y}}_{u}$.}

        \If{$\chi_u^{\left( {te} \right)}=1$}
				\State ${{\mathbf{\widehat z}}_u}^{\left( {te} \right)} = {\mathbf{\widehat z}}_{u,{\text{LoS}}}^{\left( {te} \right)}$.
                \State {Obtain ${\mathbf{\overset{\lower0.05em\hbox{$\smash{\scriptscriptstyle\smile}$}}{z} }}_{u,{\text{LoS}}}^{\left( {te} \right)}$ via loading ${\mathbf{\widehat z}}_{u,{\text{LoS}}}^{\left( {te} \right)}$ and the network }
                \State {parameter ${\mathbf{\Theta }}_{{\text{LoS-AidNet}}}$ into  LoS-AidNet}.
                \State {Obtain the detected U2G data $\widehat {\mathbf{d}}_u^{\left( {te} \right)}$ according to (\ref{ddd}) }
                \State {along with ${\mathbf{\overset{\lower0.05em\hbox{$\smash{\scriptscriptstyle\smile}$}}{z} }}_{u,{\text{LoS}}}^{\left( {te} \right)}$.}
                \State {Obtain the recovered G2U CSI ${\mathbf{\widehat h}}_u^{\left( {te} \right)}$ via loading }
                \State {${\mathbf{\overset{\lower0.05em\hbox{$\smash{\scriptscriptstyle\smile}$}}{z} }}_{u,{\text{LoS}}}^{\left( {te} \right)}$ and the network parameter ${\mathbf{\Theta }}_{{\text{CSI-RecNet}}}$}
                \State {into CSI-RecNet}.
			\Else  %\Comment{xxx}  %注释
				\State ${{\mathbf{\widehat z}}_u}^{\left( {te} \right)} = {\mathbf{\widehat z}}_{u,{\text{NLoS}}}^{\left( {te} \right)}$.
                \State {Obtain the detected U2G data $\widehat {\mathbf{d}}_u^{\left( {te} \right)}$ according to (\ref{ddd}) }
                \State {along with ${\mathbf{\widehat z}}_{u,{\text{NLoS}}}^{\left( {te} \right)}$.}
                \State {Obtain the recovered G2U CSI ${\mathbf{\widehat h}}_u^{\left( {te} \right)}$ via loading }
                \State {${\mathbf{\widehat z}}_{u,{\text{NLoS}}}^{\left( {te} \right)}$ and the network parameter ${\mathbf{\Theta }}_{{\text{CSI-RecNet}}}$}
                \State {into CSI-RecNet}.
			\EndIf

	\end{algorithmic}
%\end{algorithm}
%    \end{flushleft}
\end{breakablealgorithm}

\subsubsection*{{2) Online deployment}}
At the online running stage, with the received signal $\widetilde {\mathbf{Y}}_{u}$ (i.e., $\mathbf{Y}_{u}$ in system model at online operation stage), we employ the LS channel estimation to gain the U2G channel matrix $\widetilde {\mathbf{G}}_u^{\left( {te} \right)}$.
Meanwhile, we utilize the initial feature extraction to obtain the compressed G2U CSI initial feature ${{\mathbf{\widehat z}}_u}^{\left( {te} \right)}$ from the received signal $\widetilde {\mathbf{Y}}_{u}$.
Then, with $\widetilde {\mathbf{G}}_u^{\left( {te} \right)}$ and ${{\mathbf{\Theta }}_\textrm{LoS-SenNet}}$, we obtain the sensed result $\chi_u^{\left( {te} \right)}$ via LoS-SenNet and the hard decision operation.
If $\chi_u^{\left( {te} \right)}=1$, the G2U channel includes the LoS path, i.e., ${{\mathbf{\widehat z}}_u}^{\left( {te} \right)} = {\mathbf{\widehat z}}_{u,{\text{LoS}}}^{\left( {te} \right)}$.
If $\chi_u^{\left( {te} \right)}=0$, the G2U channel excludes the LoS path, i.e., ${{\mathbf{\widehat z}}_u}^{\left( {te} \right)} = {\mathbf{\widehat z}}_{u,{\text{NLoS}}}^{\left( {te} \right)}$.
When ${{\mathbf{\widehat z}}_u}^{\left( {te} \right)} = {\mathbf{\widehat z}}_{u,{\text{LoS}}}^{\left( {te} \right)}$, we load ${\mathbf{\widehat z}}_{u,{\text{LoS}}}^{\left( {te} \right)}$ and ${\mathbf{\Theta }}_{{\text{LoS-AidNet}}}$ into LoS-AidNet to get ${\mathbf{\overset{\lower0.05em\hbox{$\smash{\scriptscriptstyle\smile}$}}{z} }}_{u,{\text{LoS}}}^{\left( {te} \right)}$.
Then, according to (\ref{ddd}) along with ${\mathbf{\overset{\lower0.05em\hbox{$\smash{\scriptscriptstyle\smile}$}}{z} }}_{u,{\text{LoS}}}^{\left( {te} \right)}$, the detected U2G data $\widehat {\mathbf{d}}_u^{\left( {te} \right)}$ is obtained.
Simultaneously, the recovered G2U CSI ${\mathbf{\widehat h}}_u^{\left( {te} \right)}$ is obtained by CSI-RecNet with ${{\mathbf{\Theta }}_\textrm{CSI-RecNet}}$ and ${\mathbf{\overset{\lower0.05em\hbox{$\smash{\scriptscriptstyle\smile}$}}{z} }}_{u,{\text{LoS}}}^{\left( {te} \right)}$.
When ${{\mathbf{\widehat z}}_u}^{\left( {te} \right)} = {\mathbf{\widehat z}}_{u,{\text{NLoS}}}^{\left( {te} \right)}$, according to (\ref{ddd}), we directly utilize the superimposed interference cancellation method to gain the detected U2G data $\widehat {\mathbf{d}}_u^{\left( {te} \right)}$.
Meanwhile, the recovered G2U CSI ${\mathbf{\widehat h}}_u^{\left( {te} \right)}$ is obtained via loading ${\mathbf{\widehat z}}_{u,{\text{NLoS}}}^{\left( {te} \right)}$ and ${{\mathbf{\Theta }}_\textrm{CSI-RecNet}}$ into CSI-RecNet.

\section{Energy consumption and computational complexity}

%*******TableIII************
\begin{table*}[]
\renewcommand{\arraystretch}{1.2}
\caption{Computational complexity.}
\label{table_III}
\centering
\scalebox{0.95}{
\begin{tabu}{@{}c|c|c@{}}
\Xhline{1pt}
Method       & Complexity  &  Case ($N=64,M=512,L_a=5$)                            \\ \Xhline{1pt}
 Proposed     & $3N+7N^{2}+3N^{2}M+2NM+9L_aN+({\lfloor {{L_a}/3} \rfloor   \lfloor {N/3} \rfloor  })/4+2L_aN^{2}+2L_a^{2}N^{2}$  &$6,634,507$  \\ \hline
Ref\cite{x1}     & $6N+6NM+6L_aN^{2}+6L_aN^{2}M$  &$63,234,432$   \\ \hline
Ref\cite{c2} & $4L_aNM+32ML_aN+32M^{2}$  & $14,286,848$    \\ \Xhline{1pt}
\end{tabu}}

\end{table*}

\subsection*{4.1. Energy consumption analysis }

{Compared with the non-superimposed scheme (i.e., the mode of time division multiplexing) with the same transmitted power and data rate}, the proposed scheme saves the energy consumption of the UAV transmitter {due to its superimposed mode,} and thus prolongs the UAV's battery life.
For the non-superimposed scheme, {$M+N$ symbols are transmitted from UAV to gBS, where $M$ is the symbol number of U2G data and $N$ is the symbol number of G2U CSI.
By contrast, the number of the transmitted symbols in the proposed scheme is reduced to $M$. {With} transmitted power $E_u$ and symbol period $T_\mathrm{sym}$, the energy consumption of non-superimposed CSI feedback is $(M+N){E_u}T_\mathrm{sym}$, while the energy consumed by the proposed scheme is $ME_uT_\mathrm{sym}$.
To this end, the energy of $NE_uT_\mathrm{sym}$ is saved for UAV by utilizing the proposed scheme. Usually, to remedy the significant path attenuation, the gBS in UAV-assisted mmWave systems equips massive antennas{\cite{a152}}, resulting in the dimension of G2U CSI (i.e., $N$) being {prohibitively} large{\cite{a153}}.
Thus, the saved energy, i.e., $NE_uT_\mathrm{sym}$, is substantially effective in prolonging the battery life of a UAV.}

\subsection*{{4.2. Computational complexity analysis}}
The comparison of computational complexity is given in TABLE~\ref{table_III}, where the number of floating-point operations (FLOPs) is considered as the metric of computational complexity.
The number of FLOPs is obtained by counting how many computations the model does, which determines the computational time of the model{\cite{a16}}. In this paper, the complex multiplication is primarily counted as a single FLOP. {For description convenience, ``Proposed'' is used to denote the proposed LoS sensing-assisted superimposed CSI feedback; ``Ref\cite{x1}'' represents the conventional superimposed CSI feedback in \cite{x1}; ``Ref\cite{c2}'' stands for the ELM-based CSI feedback in \cite{c2}.}

In the proposed scheme, the computational complexity of LoS-SenNet is equivalent to $(36L_aN+{\lfloor {{L_a}/3} \rfloor   \lfloor {N/3} \rfloor  }) / 4$, the initial feature extraction has the computational complexity of $3N+3N^{2}+3N^{2}M+2NM$, the computational complexity of LoS-AidNet is equivalent to $(16N^{2})/4$, and the computational complexity of CSI-RecNet is equivalent to $(8L_aN^{2}+8L_a^{2}N^{2})/4$.
Thus, the total complexity of the proposed scheme (including LoS-SenNet, initial feature extraction, LoS-AidNet, and CSI-RecNet) is $3N+7N^{2}+3N^{2}M+2NM+9L_aN+({\lfloor {{L_a}/3} \rfloor   \lfloor {N/3} \rfloor  })/4+2L_aN^{2}+2L_a^{2}N^{2}$.
In addition, the total complexity of the proposed scheme without LoS-SenNet and LoS-AidNet is $3N+3N^{2}+3N^{2}M+2NM+2L_aN^{2}+2L_a^{2}N^{2}$.
The conventional superimposed CSI feedback in \cite{x1} has the computational complexity of $6N+6NM+6L_aN^{2}+6L_aN^{2}M$.
In \cite{c2}, the computational complexity is $4L_aNM+32ML_aN+32M^{2}$.
For the case where $N = 64$, $M = 512$, and $L_a=5$, the computational complexities in ``Proposed'', ``Ref\cite{x1}'', and ``Ref\cite{c2}'' are $6,634,507$, $63,234,432$, and $14,286,848$, respectively. Therefore, the proposed scheme has a lower computational complexity than those of \cite{x1} and \cite{c2}.

{On the whole, the proposed scheme saves the energy consumption of UAV transmitter and reduces the computational complexity of gBS receiver. With the saved UAV transmitter's energy consumption and the reduced gBS receiver's computational complexity, we further validate the proposed scheme can improve the normalized mean squared error (NMSE) and bit error rate (BER) at gBS {in Section 5}.}

\section{Experiment results}
{With these benefits of the UAV transmitter and gBS receiver given in Section 4,} in this section, we validate the proposed scheme can further improve the NMSE and BER of gBS receiver.
{The CDL channel model in the 5G standard{\cite{tr1}}{\cite{ts}} is employed to verify the effectiveness of the proposed scheme, which is widely used and verified to be effective{\cite{cdlreal}}}.
In Section 5.1, definitions and basic parameters involved in simulations are first given. Then, to verify the effectiveness of the proposed scheme, {the NMSE of the G2U CSI and the BER of the U2G data} are given in Section 5.2. Finally, the robustness of the proposed scheme is presented in Section {5.3.}

\subsection*{5.1. Parameters setting}
{Basic parameters and definitions} involved in simulations are given as follows. In the simulation setup, $L_a=5$, $N=64$, and $M=512$ are considered.
We adopt the Walsh matrix as the spreading matrix ${\mathbf{Q}}${\cite{y1}}.
The modulated U2G data ${\mathbf{d}}_u$ is formed by using QPSK modulation.
For LoS-SenNet, LoS-AidNet, and CSI-RecNet, the {testing set is generated with the same approach that the training set does (given in Section 3.3)}.
The definitions of NMSE and BER are followed from \cite{y1},
where the NMSE is defined as
\begin{equation}\label{NMSE}
\begin{gathered}
{\text{NMSE}} = {\text{ }}\frac{{\left\| {{{\mathbf{h}}_u} - {{\widehat {\mathbf{h}}}_u}} \right\|_2^2}}{{\left\| {{{\mathbf{h}}_u}} \right\|_2^2}}.
\end{gathered}
\end{equation}
In this paper, the NMSE and BER of the proposed scheme are compared with those of \cite{x1} and \cite{c2}.
The testing sets of LoS-SenNet, LoS-AidNet, and CSI-RecNet {contain} $30,000$, $18,000$, and $9,000$ samples, respectively.
We stop the testing for BER performance when at least 1000-bit errors are observed.
For the effectiveness validation, the power proportional coefficient is set to $\rho = 0.15$. Furthermore, the ratio of the number of G2U CSI with LoS to the total number of G2U CSI is denoted by $\beta$, which is set as $0.7$ for the validity validation.
In addition, to verify the effectiveness and robustness of LoS-SenNet and {LoS-AidNet, the} proposed scheme without LoS-SenNet and LoS-AidNet, denoted as ``Proposed (without LoS-Sen \& {LoS-Aid})'', {is also} simulated.

%=================================================FIG3
\begin{figure}[t]
\centering
\includegraphics[width=3in]{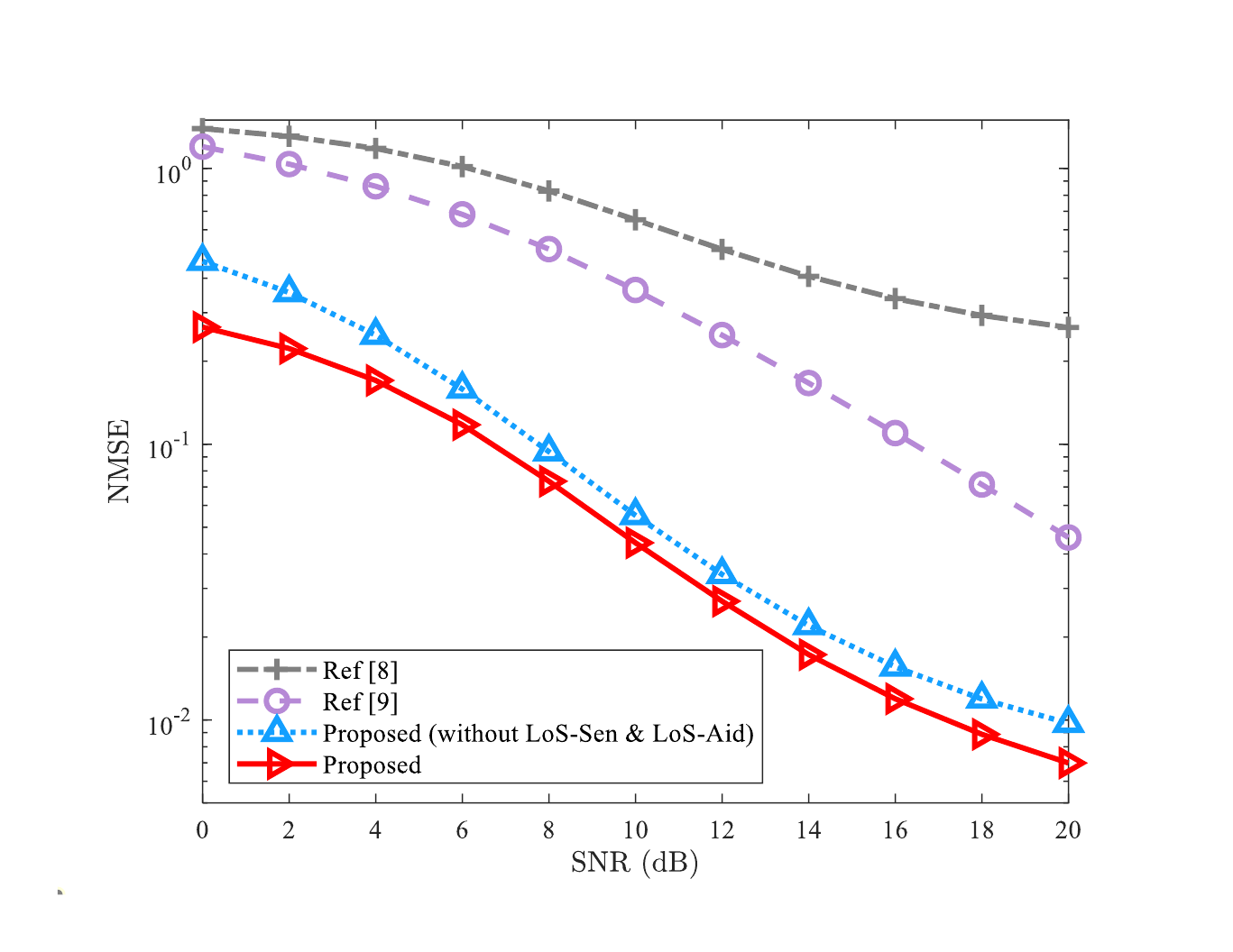}
\caption{NMSE of G2U CSI, where $L_a=5$, $N=64$, $M=512$, $\beta = 0.7$, and $\rho = 0.15$.}\label{fig2}
\end{figure}

\subsection*{5.2. NMSE and BER}
In our work, the performance tradeoff between sensing and transmission is achieved by balancing the computational complexity and the performance of NMSE and BER. To validate the effectiveness of the proposed scheme, {the NMSE of the G2U CSI and the BER of the U2G data are given in Fig.~\ref{fig2} and Fig.~\ref{fig3}, {respectively.}}

%=================================================FIG2
\begin{figure}[t]
\centering
\includegraphics[width=3in]{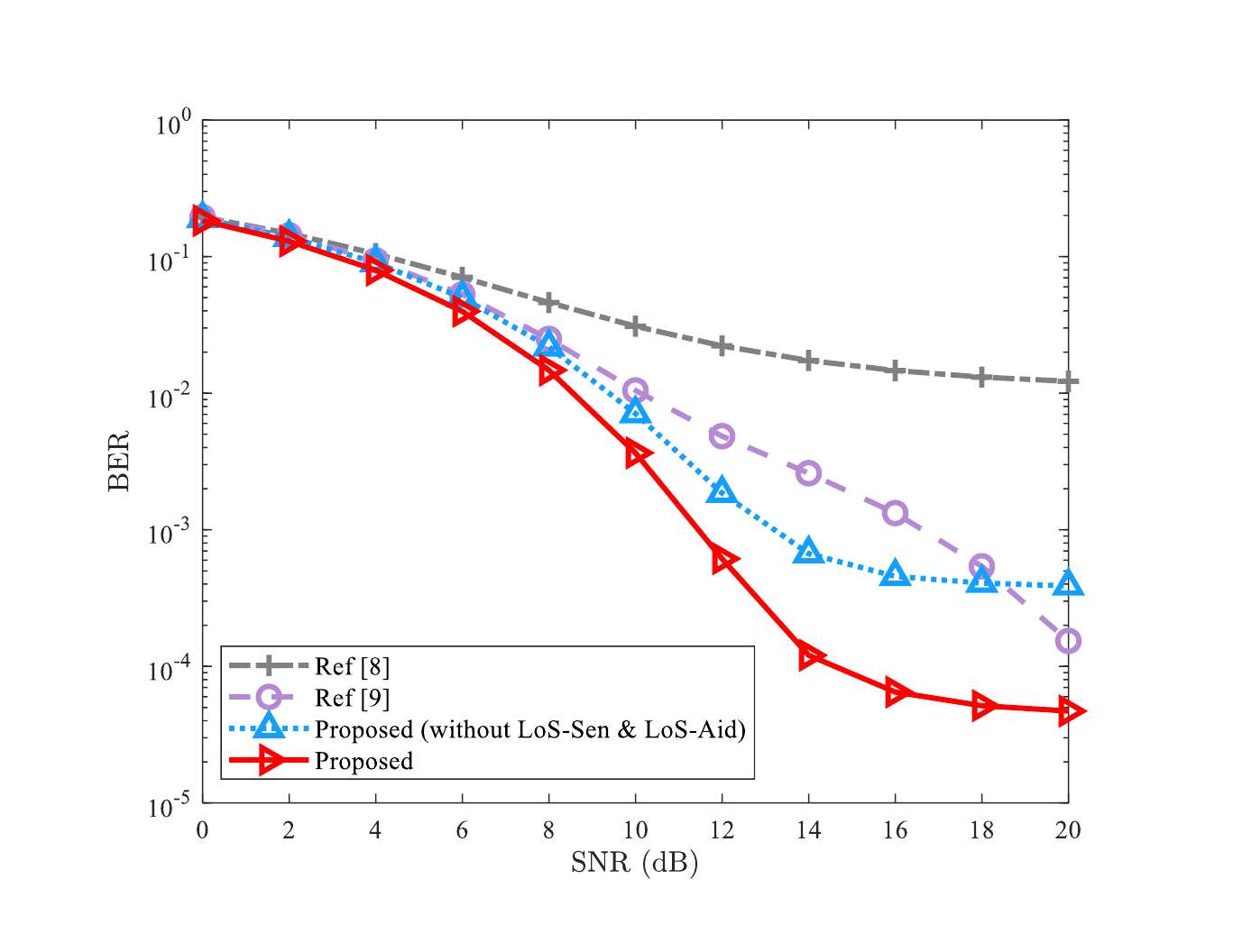}
\caption{BER of U2G data, where $L_a=5$, $N=64$, $M=512$, $\beta = 0.7$, and $\rho = 0.15$.}\label{fig3}
\end{figure}

\subsubsection*{1) NMSE performance analysis}
As shown in Fig.~\ref{fig2}, {{the NMSE curve of the ``Proposed'' is lower than} those of ``Ref\cite{x1}'' and ``Ref\cite{c2}'' for each given SNR.
For example, when ${\textrm{SNR}}=10$dB, the NMSEs of ``Ref\cite{x1}'' and ``Ref\cite{c2}'' are larger than $3.6\times 10^{-1}$, while the NMSE of ``Proposed'' is less than $4.4\times 10^{-2}$. The ``Proposed'' achieves a better NMSE performance due to the following reasons. On the one hand, at the UAV transmitter, the ``Proposed'' compresses its G2U CSI and then employs spread spectrum technology to capture a larger spread spectrum gain than those of ``Ref\cite{x1}'' and ``Ref\cite{c2}'', which effectively suppress the superimposed interference between the G2U CSI and U2G data.
Without the compression, the spread spectrum gains obtained by ``Ref\cite{x1}'' and ``Ref\cite{c2}'' are not sufficient, and thus encounter poor NMSE performance.
On the other hand, at the gBS receiver, the ``Proposed'' exploits the LoS features in the G2U CSI to assist the recovery of G2U CSI by using LoS-SenNet and LoS-AidNet.
{Furthermore, ``Proposed (without LoS-Sen \& LoS-Aid)'' is equivalent to the scheme that only considers NLoS transmission scenarios. Meanwhile, for each given SNR, the NMSE curve of ``Proposed (without LoS-Sen \& LoS-Aid)'' is lower than those of ``Ref [8]'' and ``Ref [9]'', indicating the role of CSI-RecNet in CSI reconstruction.}
In addition, from Fig.~\ref{fig2}, it can be observed that the G2U CSI's NMSE of the ``Proposed'' is smaller than that of the ``Proposed (without LoS-Sen \& LoS-Aid)'', which confirms that the {LoS-SenNet and LoS-AidNet} play an important role for the proposed scheme in the recovery of G2U CSI.
Thus, for the recovery of G2U CSI, the proposed scheme shows the smaller NMSE than those of ``Ref\cite{x1}'' and ``Ref\cite{c2}'', which reflects its effectiveness in improving the NMSE of G2U CSI. Besides, compared with the ``Proposed (without LoS-Sen \& LoS-Aid)'', {the LoS-SenNet and LoS-AidNet present their} effectiveness in reducing the NMSE of G2U CSI.}

\subsubsection*{2) BER performance analysis}
Fig.~\ref{fig3} illustrates the effectiveness of the proposed scheme in terms of the BER of U2G data.
As shown in Fig.~\ref{fig3}, the values of BER for ``Ref\cite{x1}'' and ``Ref\cite{c2}'' are considerably higher than that of ``Proposed'' {in the given SNR region.}
For example, when ${\textrm{SNR}}=10$dB, the BERs of ``Ref\cite{x1}'' and ``Ref\cite{c2}'' reach about $3.1\times 10^{-2}$ and $1.0\times 10^{-2}$, respectively, while the BER of {the} ``Proposed'' is less than $3.6\times 10^{-3}$. {The ``Proposed'' improves the BER performance compared with ``Ref\cite{x1}'' and ``Ref\cite{c2}''.
One of the reasons is because the G2U CSI is compressed at the UAV transmitter, achieving a larger spread spectrum gain than those of ``Ref\cite{x1}'' and ``Ref\cite{c2}''. In particular, at the gBS receiver,} the ``Proposed'' optimizes the recovery of compressed G2U CSI by using LoS-SenNet and LoS-AidNet to respectively sense LoS scenario and exploit the LoS features, which eases superimposed interference cancellation and thus improves the detection correctness.
{Furthermore, ``Proposed (without LoS-Sen \& LoS-Aid)'' could be regarded as the scheme that only considers the NLoS transmission scenario, and its BER curve for each given SNR is lower than that of ``Ref [8]'' and equivalent to that of ``Ref [9]'', which benefits from the spread spectrum gain at the UAV transmitter.}
In addition, the U2G data's BER of the ``Proposed'' is smaller than that of the ``Proposed (without LoS-Sen \& LoS-Aid)'', reflecting the {LoS-SenNet and LoS-AidNet} are effective to exploit the LoS features to improve the recovery of compressed G2U CSI.
On the whole, compared with the ``Ref\cite{x1}'' and ``Ref\cite{c2}'', the proposed scheme {is effective to reduce the BER of the U2G data. Relative to ``Proposed (without LoS-Sen \& LoS-Aid)'', the effectiveness of the LoS-AidNet is also validated.}

\subsection*{5.3. Robustness analysis}
In this subsection, the robustness of the ``proposed'' scheme is analysed against the impacts of varying parameters, i.e., the power proportional coefficient $\rho$ and the ratio of the number of G2U CSI with LoS to the total G2U CSI set $\beta$. {For ease of analysis}, only the {analysed parameter (i.e., $\rho$ or $\beta$) is varied} while {the} other fundamental parameters remain the same as those given in Section 5.1.

\subsubsection*{1) Robustness against $\rho$}

%=================================================FIG4
\begin{figure}[t]
\centering
\includegraphics[width=3in]{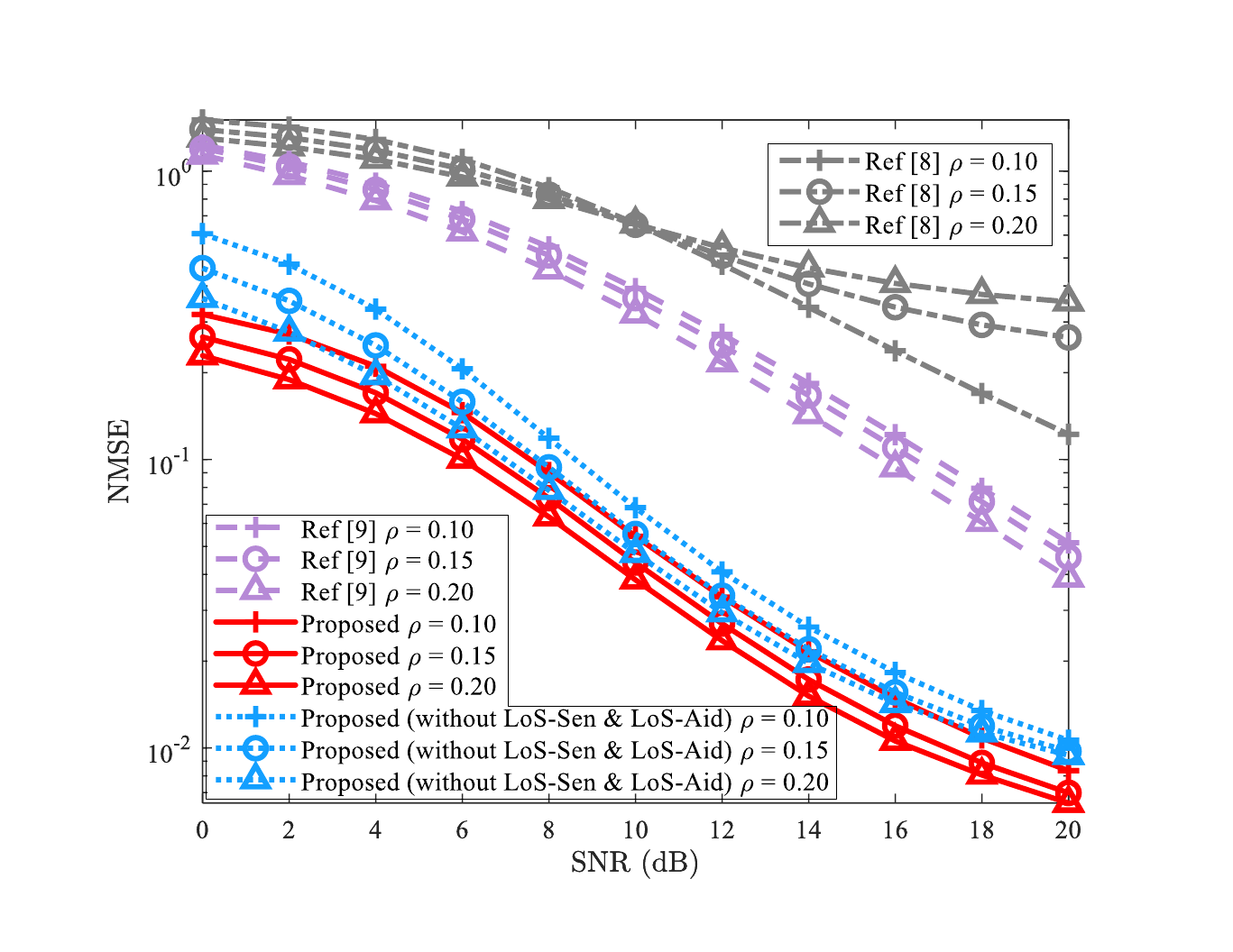}
\caption{NMSE of G2U CSI, where $L_a=5$, $N=64$, $M=512$, and $\beta = 0.7$.}\label{fig4}
\end{figure}

To evaluate the NMSE performance against the impact of $\rho$, the NMSE curves against {varying} $\rho$ (i.e., $\rho=0.10$, $\rho=0.15$, and $\rho=0.20$) are presented in Fig.~\ref{fig4}.
For each given $\rho$, the NMSE of the ``Proposed'' is smaller than those of the ``Ref\cite{x1}'' and ``Ref\cite{c2}'' {in the given SNR region.}
As the increase of $\rho$, the NMSEs decrease for ``Ref\cite{c2}'', ``Proposed (without LoS-Sen \& LoS-Aid)'', and ``Proposed'', and vice versa.
The reason is that the G2U CSI could obtain more transmission power with a larger value of $\rho$.
The NMSE of ``Ref\cite{x1}'' {decreases} and then increases with the increase of $\rho$.
{The reason is that, with relatively low SNR, e.g., SNR$\leq10$dB,} the NMSE performance is {mainly} affected by the noise and superimposed interference. {For the relatively high SNR, e.g., SNR$>10$dB, the NMSE performance is mainly affected by superposition interference.}
On the whole, for each given value of $\rho$, the NMSE of the ``Proposed'' is smaller than those of ``Ref\cite{x1}'', ``Ref\cite{c2}'', and ``Proposed (without LoS-Sen \& LoS-Aid)'' in the whole SNR regions.
For example, in the case of $\rho=0.20$ and ${\textrm{SNR}}=14$dB, the NMSE of ``Proposed'' is less than $1.5\times 10^{-2}$, the NMSEs of ``Ref\cite{x1}'', ``Ref\cite{c2}'', and ``Proposed (without LoS-Sen \& LoS-Aid)'' reach about $4.6\times 10^{-1}$, $1.4\times 10^{-1}$, and $2.0\times 10^{-2}$, respectively.
{Thus, against the impact of $\rho$, the proposed scheme reduces the NMSE of G2U CSI compared with ``Ref\cite{x1}'', ``Ref\cite{c2}'', and ``Proposed (without LoS-Sen \& LoS-Aid)''.}

%%=================================================FIG5
\begin{figure}[t]
\centering
\includegraphics[width=3in]{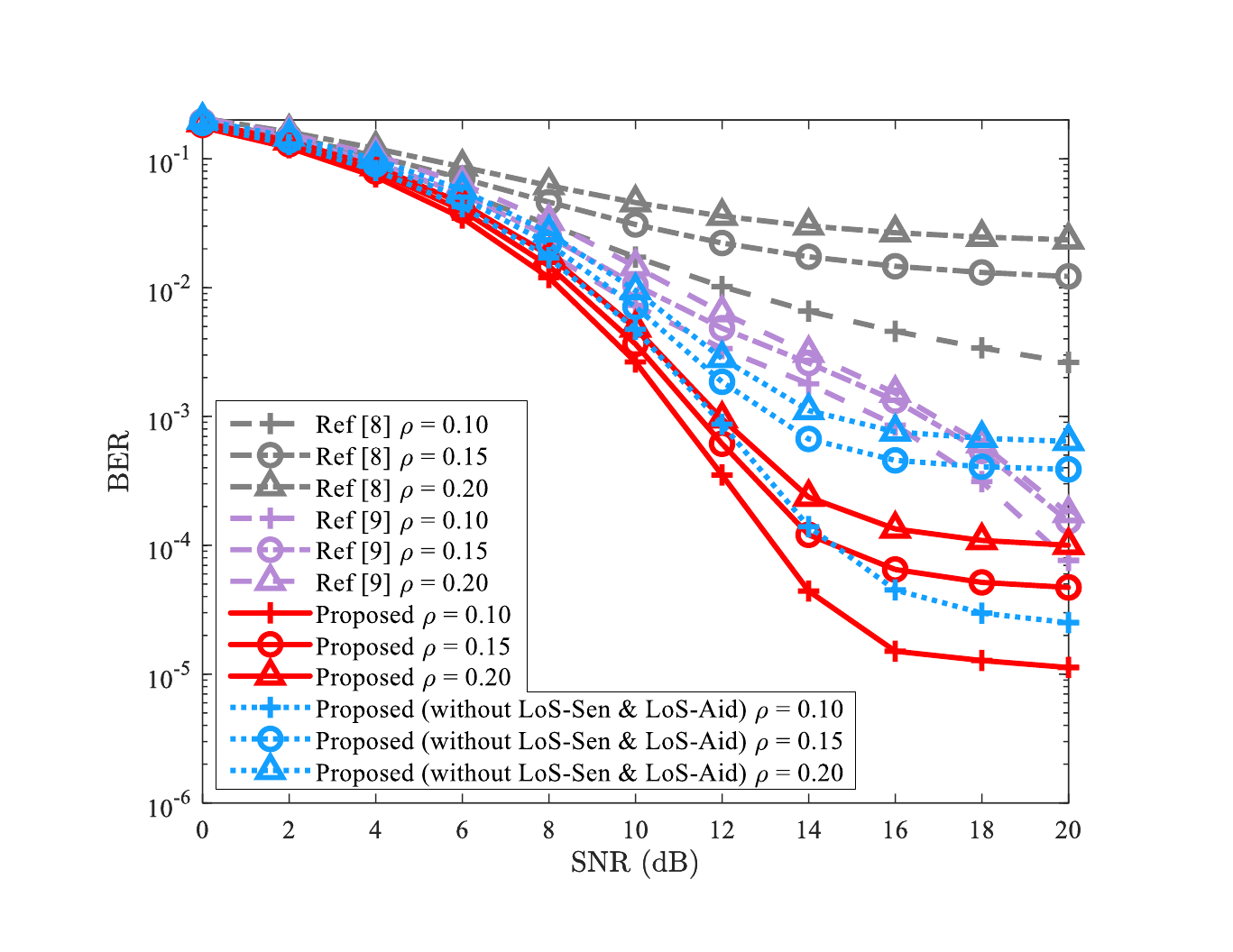}
\caption{BER of U2G data, where $L_a=5$, $N=64$, $M=512$, and $\beta = 0.7$.}\label{fig5}
\end{figure}

To verify the BER performance against the impact of $\rho$, the BER curves with {varying} $\rho$ are plotted in Fig.~\ref{fig5}.
From Fig.~\ref{fig5}, the BERs of the ``Ref\cite{x1}'', ``Ref\cite{c2}'', ``Proposed (without LoS-Sen \& LoS-Aid)'', and ``Proposed'' decrease as $\rho$ increases, and vice versa.
The reason is that the larger the power proportional coefficient $\rho$ is, the more power is allocated {to} the G2U CSI {, while} the less power is {reserved for} the U2G data, resulting in poorer U2G data detection performance.
{In addition, with the increase of SNR, the BERs of ``Ref\cite{x1}'', ``Proposed (without LoS-Sen \& LoS-Aid)'', and ``Proposed'' encounter the error floor due to the superimposed interference. Even so, for each given $\rho$ and SNR, the BER of the ``Proposed'' is smaller than those of the ``Ref\cite{x1}'', ``Ref\cite{c2}'', and ``Proposed (without LoS-Sen \& LoS-Aid)''.}
Therefore, {the proposed scheme improves the BER performance} against the impact of $\rho$.

To sum up, {against the impact of $\rho$, the proposed scheme improves the NMSE and BER compared with ``Ref\cite{x1}'', ``Ref\cite{c2}'', and ``Proposed (without LoS-Sen \& LoS-Aid)''.}

\subsubsection*{2) Robustness against $\beta$}

To provide more insights into the robustness of the NMSE performance against the impact of $\beta$, the NMSE curves {with different values of} $\beta$ (i.e., $\beta=0.60$, $\beta=0.70$, and $\beta=0.80$) are shown in Fig.~\ref{fig6}.
{From} Fig.~\ref{fig6}, for each given $\beta$, the NMSE of the ``Proposed'' is smaller than those of the ``Ref\cite{x1}'', ``Ref\cite{c2}'', and ``Proposed (without LoS-Sen \& LoS-Aid)'' in the given SNR region.
For example, in the case of $\beta=0.80$ and ${\textrm{SNR}}=16$dB, the NMSE of ``Proposed'' is less than $8.5\times 10^{-3}$, the NMSEs of ``Ref\cite{x1}'', ``Ref\cite{c2}'', and ``Proposed (without LoS-Sen \& LoS-Aid)'' reach about $3.5\times 10^{-1}$, $1.2\times 10^{-1}$, and $1.1\times 10^{-2}$, respectively.
This confirms that the proposed scheme benefits from exploiting the inherent properties of UAV-assisted mmWave systems, namely sensing the LoS scenario by LoS-SenNet and exploiting the LoS features by LoS-AidNet to assist the recovery of compressed G2U CSI.
{{As the increase} of $\beta$, the NMSEs of ``Ref\cite{x1}'' {and}  ``Ref\cite{c2}'' are {approximately} unchanged.
The reason is that the priori information of LoS is not exploited.}
On the contrary, for the ``Proposed'' {and ``Proposed (without LoS-Sen \& LoS-Aid)''}, the NMSEs decrease with the increase of $\beta$.
{The increased $\beta$ leads to a smaller sparsity due to the increased possibility of LoS, which is conducive to compressing and recovering the G2U CSI. In particular, at the gBS receiver, the LoS scenarios are sensed by LoS-SenNet and} the LoS features are further exploited by the LoS-AidNet.
{Thus, the ``Proposed'' obtains the better recovery accuracy than those of ``Ref\cite{x1}'', ``Ref\cite{c2}'', and ``Proposed (without LoS-Sen \& LoS-Aid)''.}
On the whole, against the impact of $\beta$ , the NMSEs of ``Ref\cite{x1}'', ``Ref\cite{c2}'', and ``Proposed (without LoS-Sen \& LoS-Aid)'' are reduced by using the ``Proposed''.

%=================================================FIG6
\begin{figure}[t]
\centering
\includegraphics[width=3in]{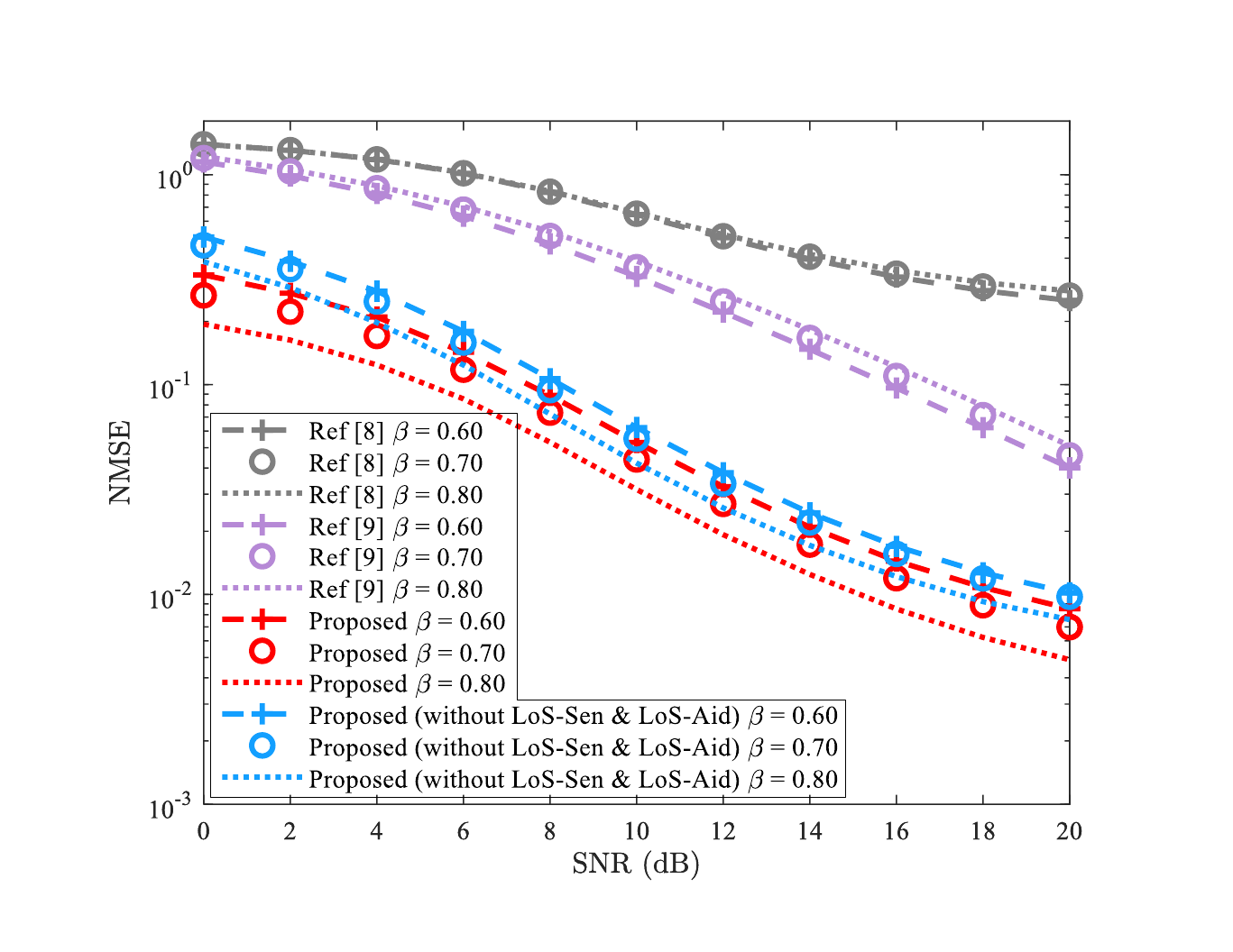}
\caption{NMSE of G2U CSI, where $L_a=5$, $N=64$, $M=512$, and $\rho = 0.15$.}\label{fig6}
\end{figure}

To highlight the robustness of the BER performance against the impact of $\beta$, Fig.~\ref{fig7} depicts the BER performance for the U2G data detection.
{It can be observed from Fig.~\ref{fig7} that as $\beta$ increases from $0.6$ to $0.8$, the BERs of ``Ref\cite{x1}'', ``Ref\cite{c2}'', and ``Proposed (without LoS-Sen \& LoS-Aid)'' are {approximatively} equivalent.}
{On the contrary, the BER of the ``Proposed'' decreases as $\beta$ increases, and this could be interpreted as the proposed scheme improving the performance of BER via refining the recovery accuracy of the compressed G2U CSI. Namely, our proposal refines the recovery accuracy of the compressed G2U CSI by making full use of the priori information of LoS via LoS-SenNet and LoS-AidNet, and then utilizes the refined compressed G2U CSI for superimposed interference cancellation to recover U2G data.
Therefore, an increase in $\beta$ has the same impact on the BER of U2G data as it does on the NMSE of G2U CSI.}
Moreover, for each given $\beta$, the BER of the ``Proposed'' is smaller than those of the ``Ref\cite{x1}'', ``Ref\cite{c2}'', and ``Proposed (without LoS-Sen \& LoS-Aid)'' {in the given SNR region}.
In a word, the proposed scheme {improves the} BER performance against the impact of $\beta$.

%=================================================FIG7
\begin{figure}[t]
\centering
\includegraphics[width=3in]{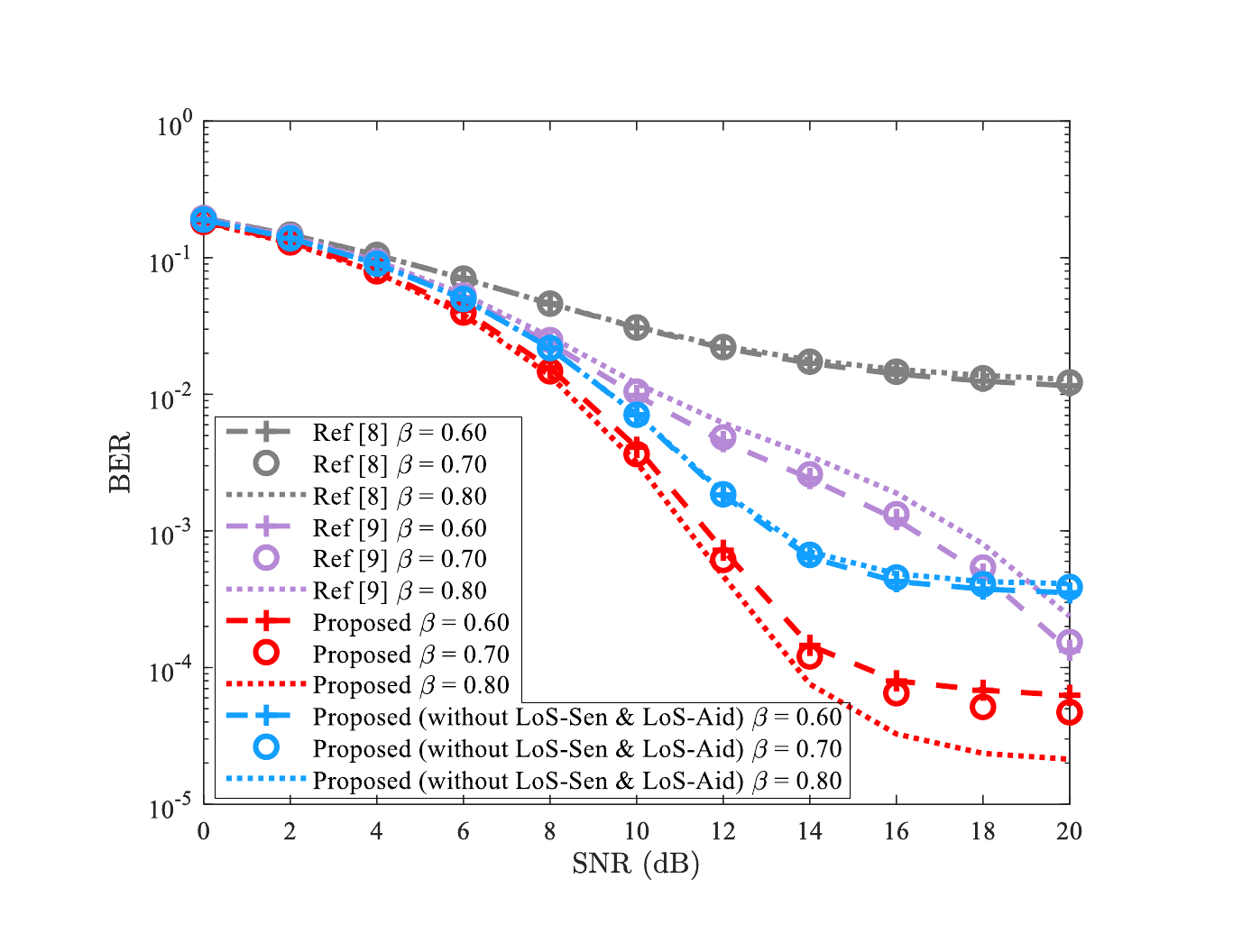}
\caption{BER of U2G data, where $L_a=5$, $N=64$, $M=512$, and $\rho = 0.15$.}\label{fig7}
\end{figure}

To sum up, {despite the impact of $\beta$, Fig.~\ref{fig6} and Fig.~\ref{fig7} illustrate that the ``Proposed'' could still improve the NMSE and BER compared with ``Ref\cite{x1}'', ``Ref\cite{c2}'', and ``Proposed (without LoS-Sen \& LoS-Aid)''.}
Furthermore, according to the curves of ``Proposed'' and ``Proposed (without LoS-Sen \& LoS-Aid)'' in Fig.~\ref{fig6} and Fig.~\ref{fig7}, it is also validated that the proposed scheme applies not only to LoS transmissions but also to NLoS transmissions.

\section{Conclusions}

{In this paper, a LoS sensing-assisted superimposed CSI feedback scheme is proposed to alleviate the significant CSI feedback overhead and high energy consumption in UAV-assisted mmWave systems.
Three lightweight networks are implemented to refine the recovery accuracy of the G2U CSI, {thereby improving the detection accuracy of the U2G data.}
{Inspired by ISAC, the first network LoS-SenNet will sense whether the U2G channel contains the LoS path, then the superimposed U2G data and G2U CSI will be recovered by the superimposed interference cancellation and two dedicated lightweight neural networks, i.e., LoS-AidNet and CSI-RecNet.}
Compared with other CSI feedback solutions, the proposed scheme could reduce the computational complexity of the gBS receiver, decrease the energy consumption of the UAV transmitter, and prolong the battery life.
Simulation results validate the effectiveness of the proposed scheme in terms of lower NMSE and BER and the robustness against the impact of variant parameters.}
{{In our future works,} we will consider the real data in real channel scenarios to promote the application of LoS sensing-based superimposed CSI feedback in practical systems.
Meanwhile, we will study online training to alleviate the impact of off-line training on real-time performance.}

\section*{Declaration of Competing Interest}
The authors declare that they have no known competing financial interests or personal relationships that could have appeared to influence the work reported in this paper.

\section*{Acknowledgements}
The authors would like to acknowledge the support of the Sichuan Science and Technology Program (Grant No.2021JDRC0003, 23ZDYF0243, 2021YFG0064), the Demonstration Project of Chengdu Major Science and Technology Application (Grant No. 2020-YF09-00048-SN), the Special Funds of Industry Development of Sichuan Province (Grant No. zyf-2018-056),
and the Industry-University Research Innovation Fund of China University (Grant No. 2021ITA10016/cxy0743).

\end{document}